\documentclass[
 aps, pra,
 amsmath,amssymb,
 11pt,
 final,
tightenlines,
 twoside,
 onecolumn,
 nofloats,
nofootinbib,
 superscriptaddress,
showkeys,
showkeywords,
 ]
{revtex4-2}

\usepackage[T2A]{fontenc}
\usepackage[utf8x]{inputenc}
\usepackage[russian,english]{babel}
\usepackage{graphicx}
\usepackage{dcolumn}
\usepackage{bm}

\setcitestyle{authoryear,round}
\def\squareforqed{\hbox{\rlap{$\sqcap$}$\sqcup$}}

\def\sq{\ifmmode\squareforqed\else{\unskip\nobreak\hfil
\penalty50\hskip1em\null\nobreak\hfil\squareforqed
\parfillskip=0pt\finalhyphendemerits=0\endgraf}\fi}

\def\utw{\smash{\rlap{\lower5pt\hbox{$\sim$}}}}

\def\udtw{\smash{\rlap{\lower6pt\hbox{$\approx$}}}}

\def\diameter{{\ifmmode\mathchoice
{\ooalign{\hfil\hbox{$\displaystyle/$}\hfil\crcr
{\hbox{$\displaystyle\mathchar"20D$}}}}
{\ooalign{\hfil\hbox{$\textstyle/$}\hfil\crcr
{\hbox{$\textstyle\mathchar"20D$}}}}
{\ooalign{\hfil\hbox{$\scriptstyle/$}\hfil\crcr
{\hbox{$\scriptstyle\mathchar"20D$}}}}
{\ooalign{\hfil\hbox{$\scriptscriptstyle/$}\hfil\crcr
{\hbox{$\scriptscriptstyle\mathchar"20D$}}}}
\else{\ooalign{\hfil/\hfil\crcr\mathhexbox20D}}%
\fi}}

\begin{document}

\selectlanguage{english}



\title{Calibration of Uncertainties of the Gaia DR3 Catalog Based on Data on Wide Binary Stars of the Galaxy Field}

\author{\firstname{D.~A.}~\surname{Kovaleva}}
 \affiliation{Institute of Astronomy of the Russian Academy of Sciences, Moscow, Russia}

\begin{abstract}
   
   {
The catalog of wide binary stars \cite{2021MNRAS.506.2269E}, created on the basis of Gaia EDR3 data and including more than a million pairs, was used to analyze Gaia DR3 data obtained independently for their components. It is shown that the spatial heterogeneity of the catalog reflects the Gaia scanning law. The change in the spatial density of binary stars in the catalog with increasing distance from the Sun has been studied. By comparison with the model distribution, it is shown that the catalog contains approximately 2.5 times fewer binary stars than would be expected in the absence of spatial incompleteness. It is confirmed that the radius of spatial completeness of the catalog is on average close to 200 pc and depends on the absolute magnitude of the main component. The spatial density of binary stars in the catalog depends weakly on the difference in the magnitudes of the components, and significantly depends on the physical distance between the components. The incompleteness of the catalog in relation to pairs with a distance between components less than 100 AU occurs already at a distance of 25 pc from the Sun. Comparison of the characteristics of components of the same pair independently determined within the Gaia DR3 catalog allowed us to study how the probability of a non-random combination of components is related to the similarity of their characteristics. A high correlation between the degree of agreement between the characteristics and the reliability of the pair was found for radial velocities. Qualitative agreement is observed for metallicity [Fe/H] estimates and, to a lesser extent, for absorption estimates. No agreement was found for the ages of the stars, which indicates their great uncertainty in the ensemble, consisting mainly of main sequence stars. At the same time, age estimates for pairs with evolved components show significantly better agreement than for the dataset as a whole. Using the parameters of the components of the pairs from Gaia DR3, an independent estimate of the uncertainties in the radial velocities and metallicities depending on the apparent magnitude of the sources was performed. Estimates of the probable median values of errors in the radial velocities and metallicities of Gaia DR3 sources are proposed. Depending on the apparent magnitude, they exceed the median error values given in the catalog: for radial velocities by 1.5–3 times, for metallicities [Fe/H] by 7–25 times. 
}

\end{abstract}

\maketitle

\section{Introduction}
It is known that the occurrence of binary and multiple stars in the Galaxy, as well as their properties, are determined by the characteristics of the star formation process and, to a certain extent, by the dynamic interaction of stars in groups with a common origin \citep{2013ARA&A..51..269D, 2012A&A...543A...8M, 2015ASPC..496...37B}. Distinguishing between the contributions of the characteristics of the emerging population and their      and non-trivial task (see, for example, the discussion in [5, 6]). The statistical properties of binary stars, obtained from observations, make it possible to introduce restrictions when solving problems related to the
formation and evolution of the stellar population of
the Galaxy \cite[see, for example, works by ][]{1982Ap&SS..88...55P, 1988Ap&SS.142..245V, 2012A&A...543A...8M,  2022A&A...659A..96M, rozner2023born}. 

The publication of each new release of data from the Gaia space project  \citep{2016A&A...595A...1G} initiates the production of new results in galactic astronomy and related fields of astronomy and astrophysics. This also applies to the study of the population of binary stars in the Galaxy. Thus, Gaia data has opened up the possibility of identifying and studying a population of wide binaries that is practically undistorted by observational selection effects. Studies carried out by different authors  independently demonstrated, in particular, a striking feature of the population of binary field stars —-- the presence of the so-called “twin peak” in the distribution of component mass ratios $q=M_{min}/M_{max}$ \citep{Badry, our}. Previously, such a peak (an excessive, in respect with the smooth distribution, of a number of binary stars with similar component masses) was detected for spectroscopic binary stars \citep{2000A&A...360..997T, 2017ApJS..230...15M}, and was presumably explained by competitive accretion onto a close binary protostar surrounded by a common envelope or accretion disk. The reliable existence of a “twin peak” for binary stars with distances between components of up
to thousands of astronomical units forces discussion of possible mechanisms for increasing the distance between components after their masses became equal \citep{Badry}. This probably occurred during dynamic interactions in the host star cluster before its disintegration and the movement of the twin stars into the field. Moreover, the distance between the components of such pairs imposes restrictions on the conditions in the cluster.

Using refined astrometric observations from the
Gaia Early Data Release 3 \citep{2021A&A...649A...1G}, \cite{2021MNRAS.506.2269E} published a list of likely members of wide binary stars within 1 kpc of the Sun, containing over a million entries. The pairs were identified as a result of the analysis of spatial and kinematic information about the probable components, considering the probability of a random coincidence in the direction of motion of the stars. In June 2022, the full Gaia Data Release 3  was published, containing, in particular, average radial velocities for 33 million stars, as well as an extensive catalog of astrophysical characteristics estimates made for large samples of stars using low- and medium-resolution Gaia spectra. In addition, a catalog of the so-called “non-single” Gaia sources was published, the solution for the motion of which assumes the presence of orbital motion due to the presence of an invisible satellite. Due to its pilot nature, the list of non-single stars has a complex selectivity function \citep{2022gdr3.reptE...7P}. Moreover, it was previously shown that the possibilities of studying binary stars with Gaia with an angular distance close to the Gaia resolution limit, as well as binaries with pronounced orbital motion, are limited \citep{Kovaleva2021125, 2022MNRAS.517.2925C}. 

In this paper, we use Gaia DR3 data to explore and independently test a catalog of probable wide binary stars \citep{2021MNRAS.506.2269E}.

In Section 2, we discuss the characteristics of the wide binary star catalog and its complement with Gaia DR3 data. Section 3 compares the characteristics of the pair components published under Gaia DR3 and examines the spatial completeness of the wide pair catalog. Section 4 discusses the use of independently determined pair component parameters to independently estimate radial velocity and metallicity uncertainties in Gaia DR3. Section 5 contains suggested conclusions.
\section{Data}
\label{data}
The Wide Binary Star Catalog \cite{2021MNRAS.506.2269E} is the largest catalog of probable binary stars and is homogeneous and well described. The catalog contains information about pairs of stars located at distances up to 1 kpc from the Sun with mutual separations from 10 AU up
to 1 pc, and having consistent proper motions. In this case, the tangential component of the relative velocity of the components and the distance between the components are such that they imply a low probability of a random coincidence of velocities. The authors of the catalog chose the {\it R\_chance\_alignment} metric as a numerical characteristic of this probability, and for pairs considered as real, $R\_chance\_alignment \leq 0.1$. This condition is satisfied by 1023441 pairs of stars out of 1571545 pairs in the catalog. 

Let us discuss the selectivity function of catalog \cite{2021MNRAS.506.2269E} determined by the method of its creation. It is formed by the selectivity of the Gaia EDR3 source catalog, including its resolution, which depends on the brightness difference of the sources $\Delta G$ (approximately, it can be stated that sources, regardless of the brightness difference $\Delta G$, are resolved at angular distances wider than $2^{''}$; at distances from $1^{''}$, sources with a brightness difference $\Delta G<4^m$ are resolved). The boundary of the incompleteness region is blurred. This effect disrupts the
completeness of the binary sample at distances up to about 200 pc from the Sun. It can be shown that the boundary of the region of incompleteness due to Gaia’s resolution depends on the scanning law, and not on the density of the stellar field, as one would expect. In the left panel, Fig.~1 shows the distribution on the celestial sphere of the minimum angular distance $\theta_{min}$ (in arcseconds) between the components of the pair depending on the galactic coordinates {\it (l, b)} of the first component. Traces of the preferred scanning directions are clearly visible. A similar effect remains for the tangential projection of the minimum linear distance between components (right panel of Fig. 1). We can confidently expect this effect to decrease with subsequent releases of Gaia data. It should be noted that this dependence is not related to the distribution of catalog sources over the celestial sphere, which reflects the nature of the distribution of stars in the Galaxy \citep[see Fig.~1 in the article]{2021MNRAS.506.2269E}.


   \begin{figure}
     \centering
         \centering
         \includegraphics[width=8.9cm]{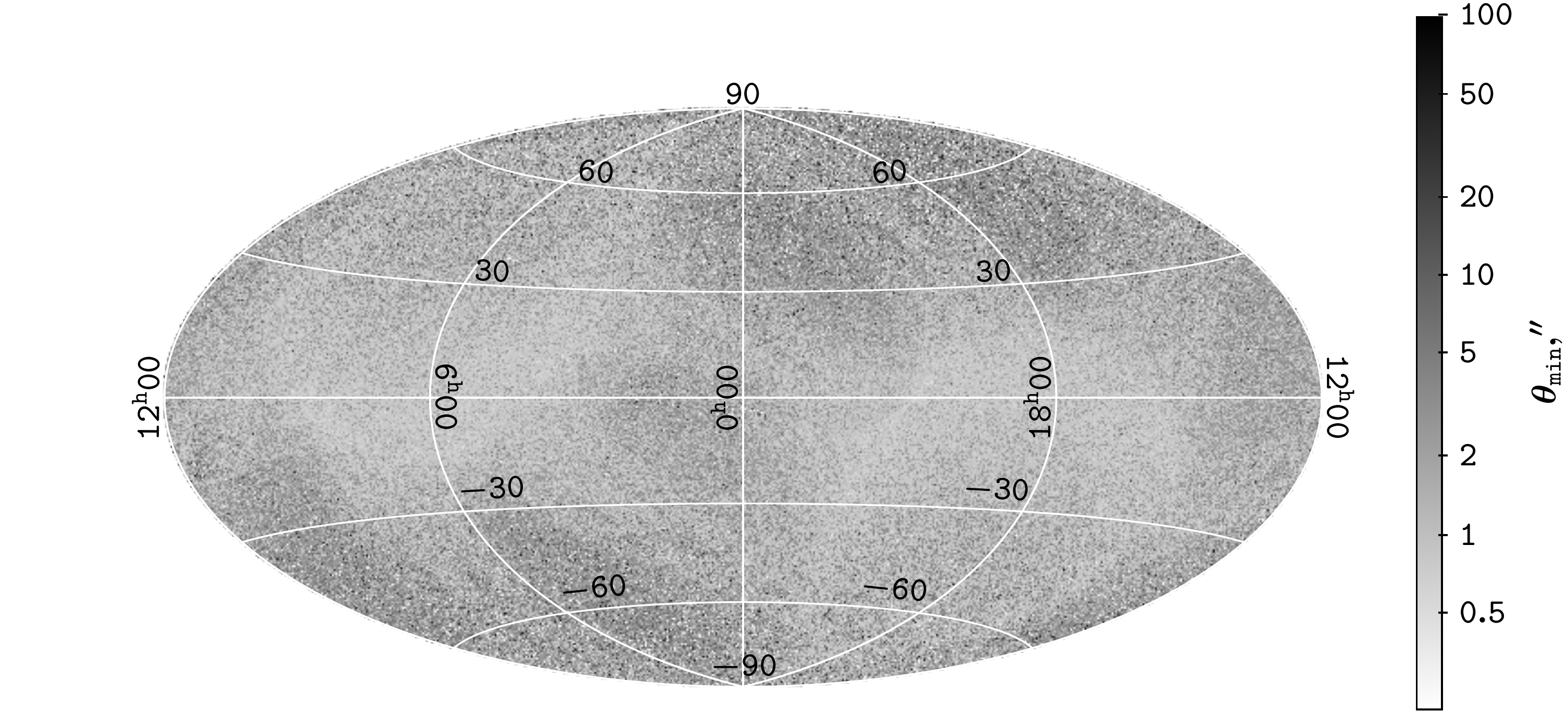}
     \hfill
         \centering
         \includegraphics[width=7.4cm]{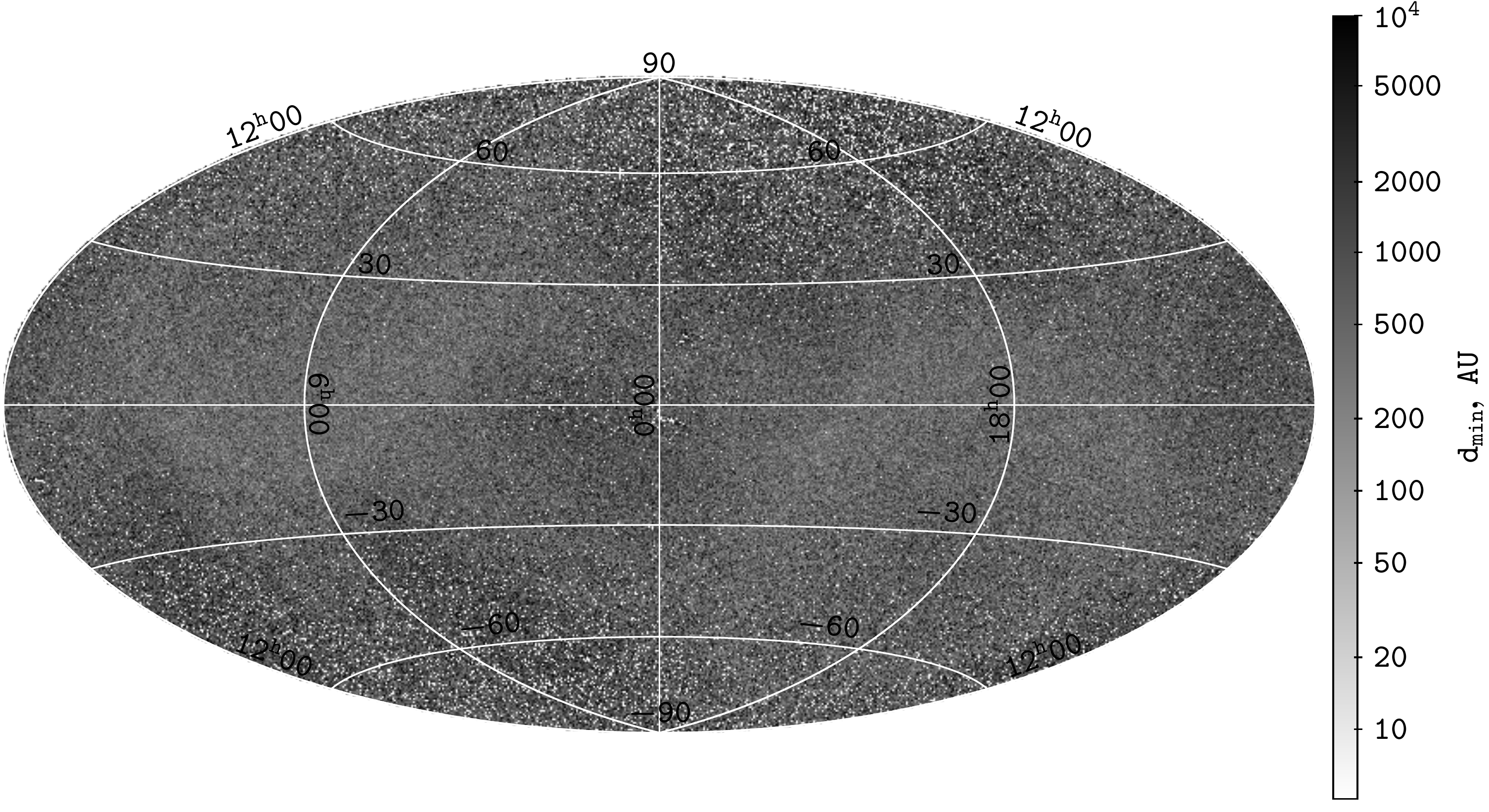}
  \caption{Dependence of angular resolution on the Gaia scanning law. Left panel shows the minimum angular separation between
the components of wide pairs of the catalog \cite{2021MNRAS.506.2269E} depending on the galactic coordinates {\it (l, b)}. The right panel is the same, but for
the minimum value of the tangential projection of the linear distance between the components.}
  
     \label{FigAngDist}
     \end{figure}

At longer distances, a different selection effect operates: pairs, one of both components of which are weaker than the limiting magnitude of Gaia, disappear from consideration. And for this area of incompleteness the boundary is not clear; for different areas of the sky the limiting value $G_{lim}$ can vary within $19-21^m$. When creating the catalog, the authors did not use filters based on the quality of the astrometric and/or photometric solution (except for the limitation on the relative error of trigonometric parallax), as these filters would have effectively filter out stars with pronounced orbital motion. However, deterioration in the quality of the astrometric solution for such stars could in itself, in some cases, lead to distortions in the values of astrometric parameters and the failure to detect a probable pair. An indirect indication of such possibility is the underestimation of the nominal parallax errors of close components discovered for catalog
stars. In addition to the effects mentioned, when creating the catalog, its authors deliberately excluded from the catalog sources in resolved triple systems, moving groups and star clusters.

Thus, the completeness of the sample of binary stars presented in the catalog \cite{2021MNRAS.506.2269E} in parameter space is limited and affected by a number of factors. However, it has a high degree of homogeneity and is the most representative of the known lists of binary stars, making it an attractive object for study. 

When creating the catalog \cite{2021MNRAS.506.2269E}, information about the radial velocities of stars —- probable components of pairs —- was not considered in any way. Radial velocities for more than 33 million sources \citep{2022arXiv220605902K} are published as part of the main Gaia DR3 catalog; for 122003 pairs from the catalog under study, radial velocities are available for both components. We match the Gaia
DR3 radial velocity information for the probable components of the pairs. The average nominal error in determining the radial velocities of individual sources for an ensemble of catalog stars is 2.6 km/s.

Within Gaia DR3, estimates of the astrophysical characteristics of Gaia sources were carried out \citep{2022arXiv220605864C, 2022AAS...24033303F}, based on a combination of astrometric and photometeric data with low- and, in certain cases, medium-resolution spectroscopy with models of stellar evolution, considering the model of
the distribution of stars in the Galaxy. In this case, a set of astrophysical characteristics for each star was determined simultaneously in such a way as to best satisfy the observational and a priori specified parameters. The catalog of wide pairs makes it possible to externally assess the accuracy of the obtained astrophysical parameters under the assumption that the components have the same origin, similar ages, contents of chemical elements, and the interstellar absorption for them should be similar. For 161944 pairs from the catalog, age estimates are available for
both components; for 666903 pairs, metallicity [Fe/H] and interstellar extinction in the G-band $A_G$ estimates are available. 
These estimates were obtained without using medium resolution spectroscopy. Within Gaia DR3, the most probable values and the lower and upper limits of the range of values corresponding to the first and third quartiles of the probability distribution are given for astrophysical characteristics. The probable range of values is asymmetric relative to the point estimate, but
if we use half the difference between the largest and smallest of the boundary values as a characteristic error, then the average accuracy of age determination for catalog sources of wide pairs is $\epsilon(Age) \approx 3.8$ million years; for metallicity estimates the average specified accuracy is $\epsilon [Fe/H] \approx 0.1$, and for extinction $\epsilon (A_G) \approx 0.2$. Metallicity values [Fe/H]$_S$ obtained based on the data from Gaia spectrograph are available for both components of just 2400 pairs. We will not discuss them within the scope of this work.

In further analysis, we also use geometric estimates of distances to stars in the catalog \cite{2021MNRAS.506.2269E} from the Catalog of probable distances to stars of Gaia EDR3 \cite{2021AJ....161..147B}.

\section{Results}
\label{res}

\subsection{Comparison of Component Characteristics}

The agreement between independently determined
characteristics of the probable components of the pairs
can be interpreted from two points of view. The first of
these is testing the hypothesis of true binarity. The
second points of view is testing assessments of independently defined characteristics. In both cases, it is
assumed that the radial velocities $RV$, estimates of
ages $Age$, metallicity [Fe/H], and extinction $A_G$ of the
components of the binary star should coincide. Obviously, such an assumption may be incorrect, at least
with respect to radial velocities, if there is orbital
motion in the system. The maximum probable contribution of orbital motion to the radial velocities of the components of the catalog \cite{2021MNRAS.506.2269E} can be several times
greater than the average radial velocity error and
amount to tens of km/s. Such an estimate can be
obtained by selecting from a list of stars with known
orbits those that should be included in the catalog of
wide pairs according to formal criteria (angular distance between the components and distance to the
Sun), and calculating their average orbital speed.
Using the catalog of astrometric orbits of binary stars
\citep{2001AJ....122.3472H}, supplemented with data from Gaia EDR3 \citep{2022MNRAS.517.2925C},
we found that for 603 binary stars with known orbits
that meet the selection criteria of the catalog of wide
pairs, the estimated average orbital velocities range
from 0.1 to 80 km/s with a median value of 2.8 km/s.

\begin{figure}
     \centering
         \centering
         \includegraphics[width=8cm]{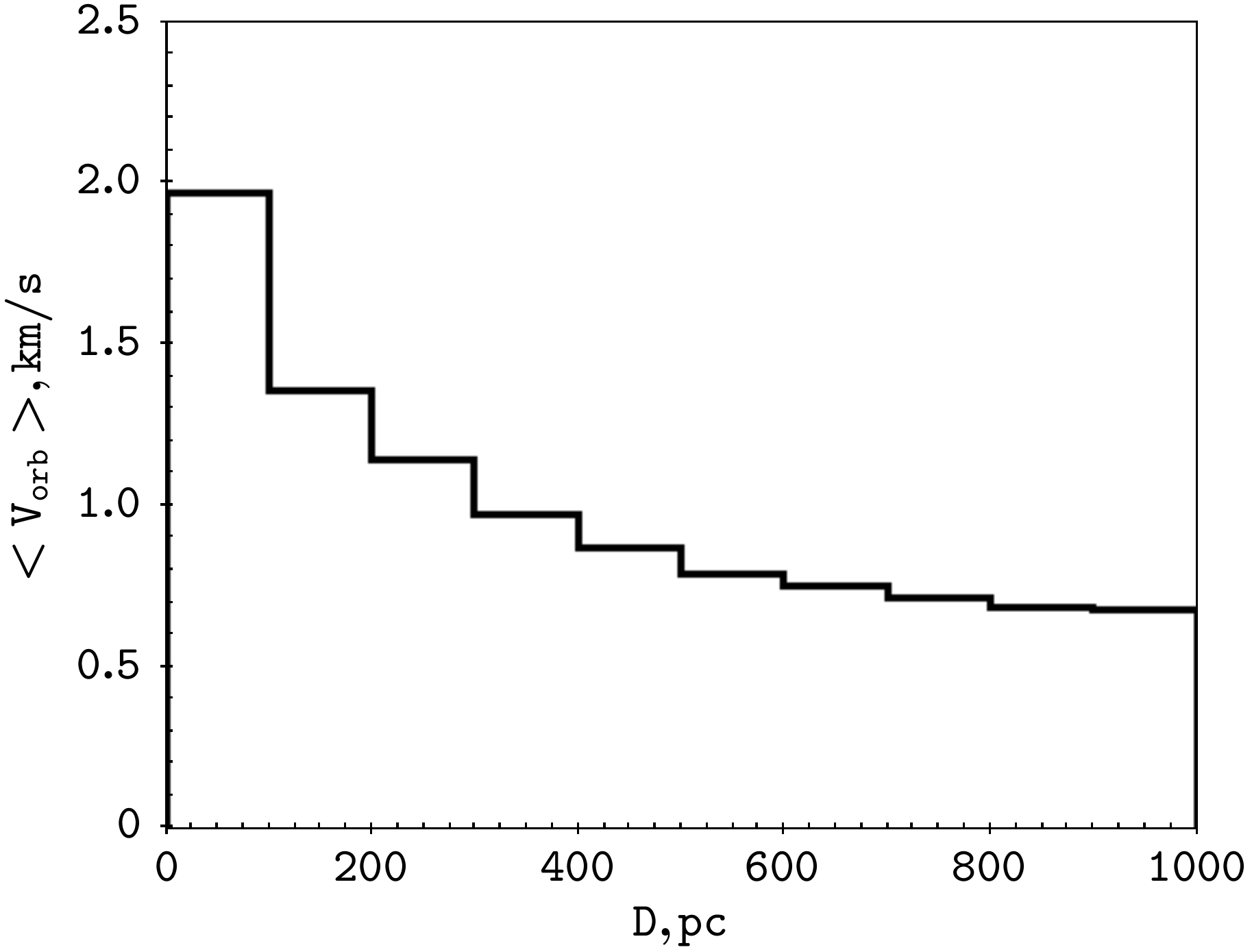}
     \hfill
         \centering
         \includegraphics[width=8cm]{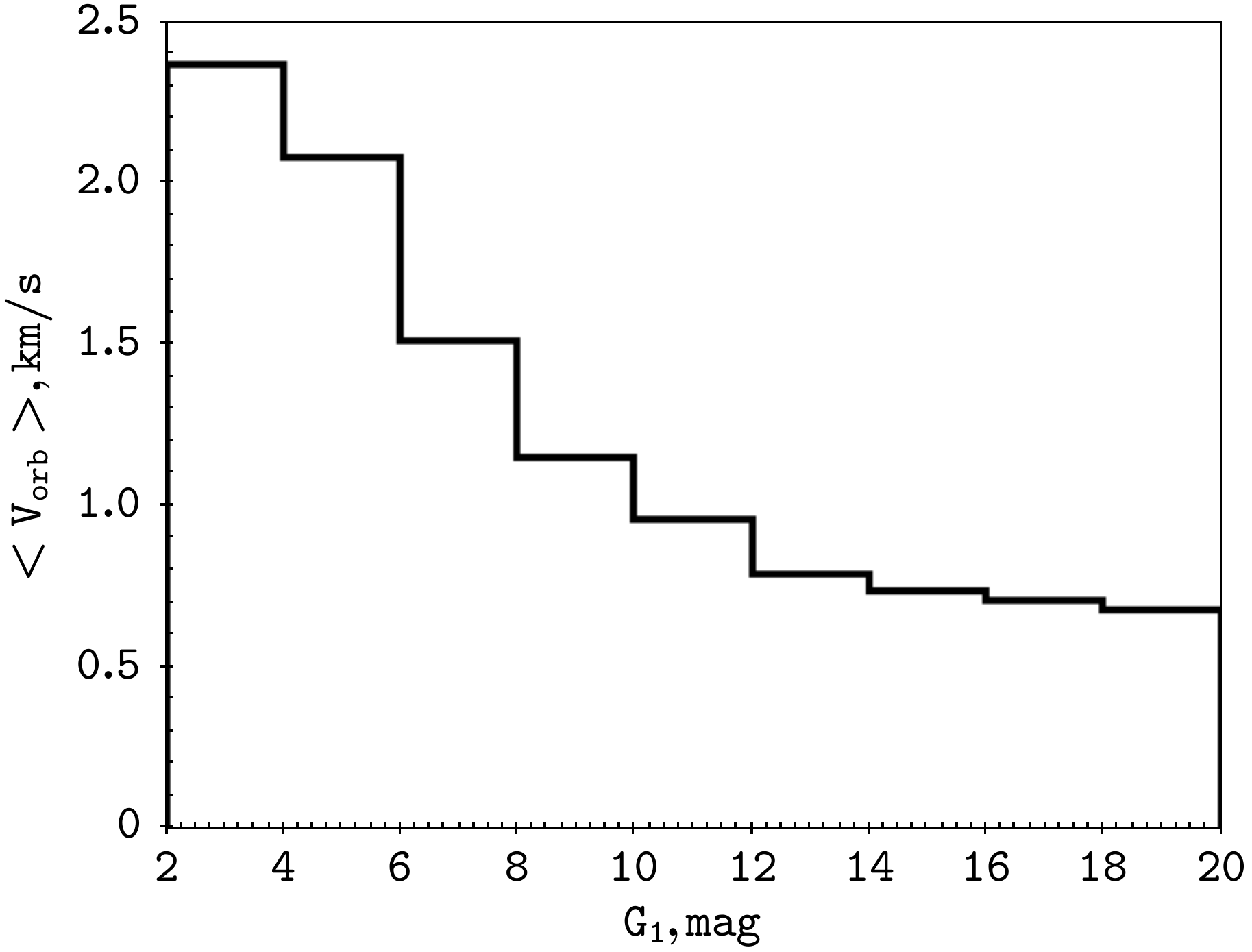}
  
  \caption{Relation of the median expected orbital speed with the distance to the main component of the system (left panel) and
with the visible magnitude of the main component (right panel).}
  
     \label{FigVmeanorb}
     \end{figure}

For 178453 pairs in the catalog, estimates of the
masses of both components are available, and the
average orbital speed can be estimated by determining
the pair’s orbital period based on Kepler’s third law.
Assuming a random orientation of circular orbits in
space, the average projection of the physical separation
between the components onto the tangential plane
is $0.8a$, where $a$ is the semimajor axis of the
orbit. The estimates of average velocities obtained in
this way range from 0 to 8 km/s, their median value is
0.7 km/s, and for $99\%$ of pairs the obtained velocities
are less than 2.7 km/s —-- which is a value close to the average
nominal error in determining the radial velocity in the
catalog. In this case, the relation of the probable
orbital speed with the distance to the system and, in
connection with this, with the apparent magnitude of
the main component is observed: at smaller distances
from the Sun, closer systems with a shorter orbital
period can be resolved. Figure~\ref{FigVmeanorb} shows the nature of
these dependencies. The histogram bars represent the
median expected orbital velocity for pairs at different
ranges of distances to the Sun and with different
apparent magnitudes of the main component. This median value varies from about 2~km/s for the closest
and brightest pairs (closer than 100 pc and/or with a
main component magnitude $G<6^m$) to less than
1~km/s for pairs further than 300~pc and/or with a
main component magnitude $G>10^m$.

\begin{figure}
     \centering
         \centering
         \includegraphics[width=8cm]{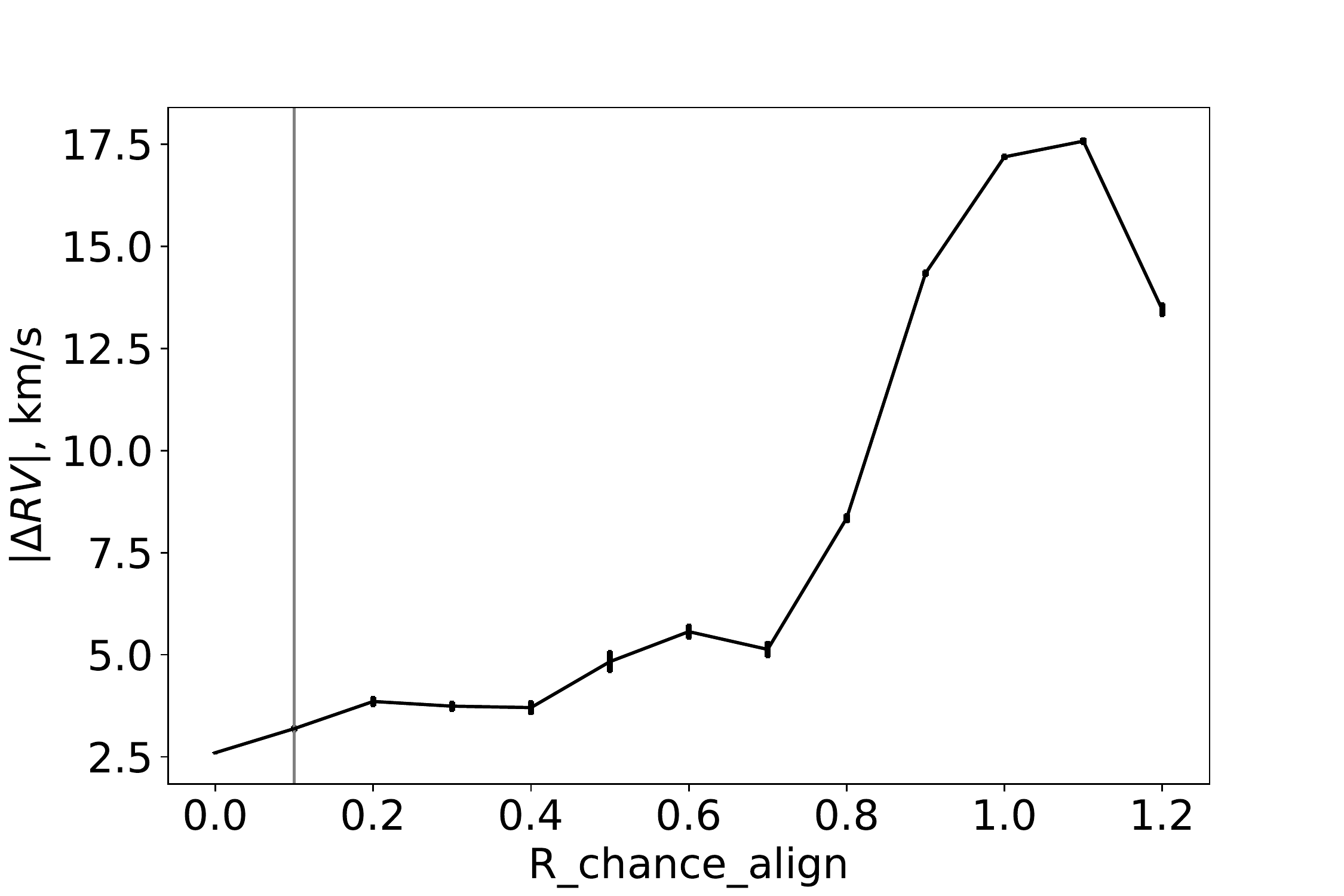}
     \hfill
         \centering
         \includegraphics[width=8cm]{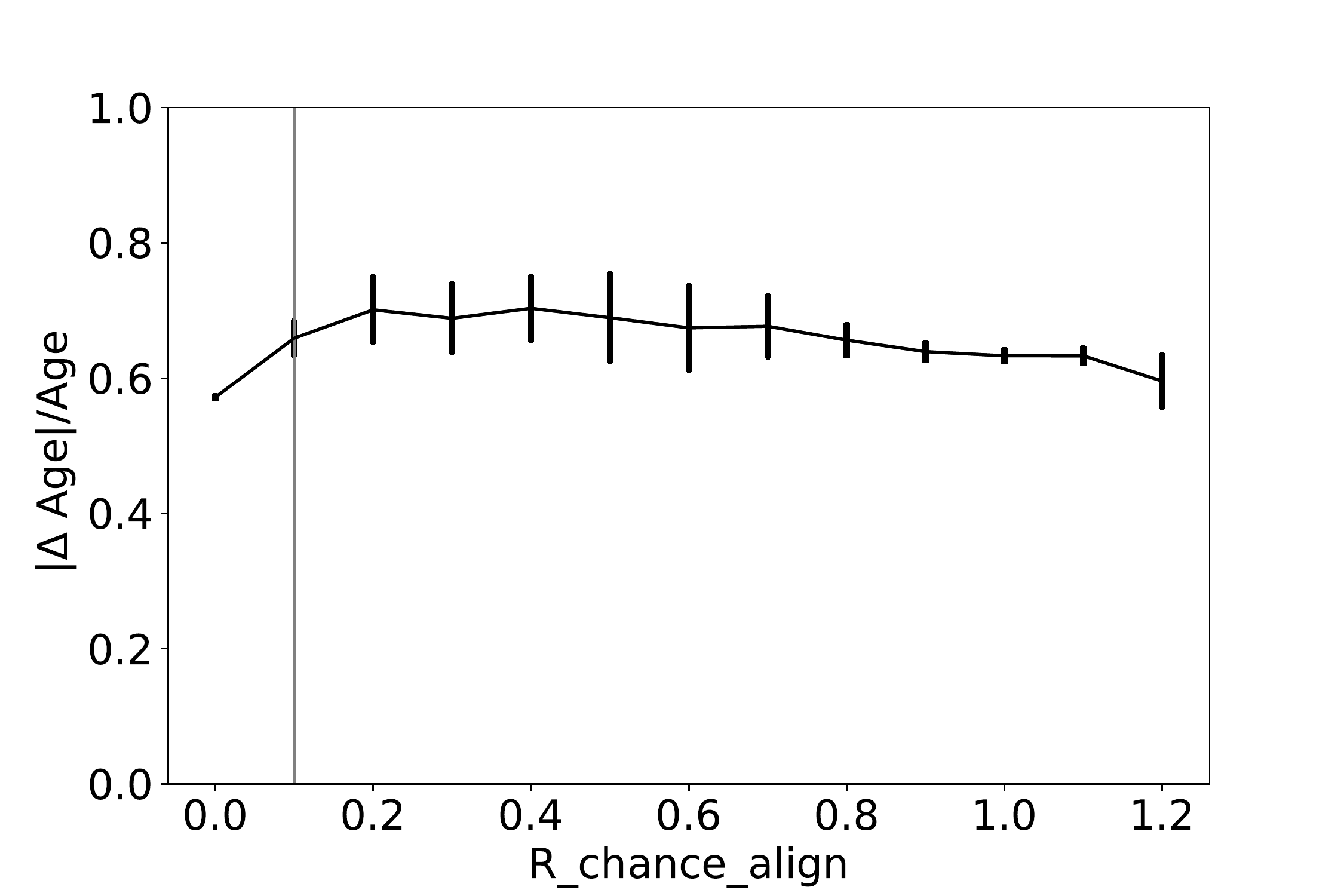}
    \hfill
         \centering
         \includegraphics[width=8cm]{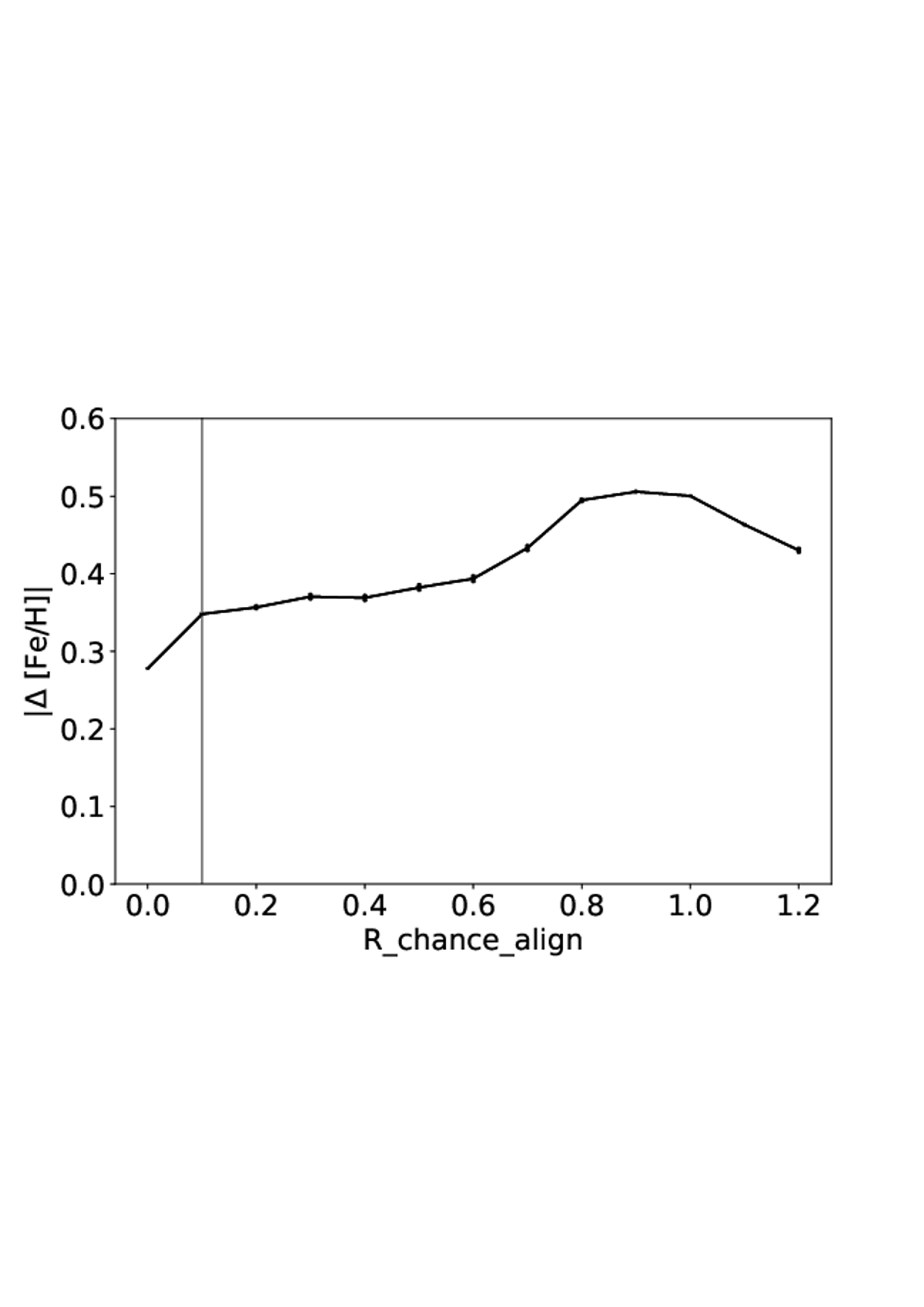}
     \hfill
         \centering
         \includegraphics[width=8cm]{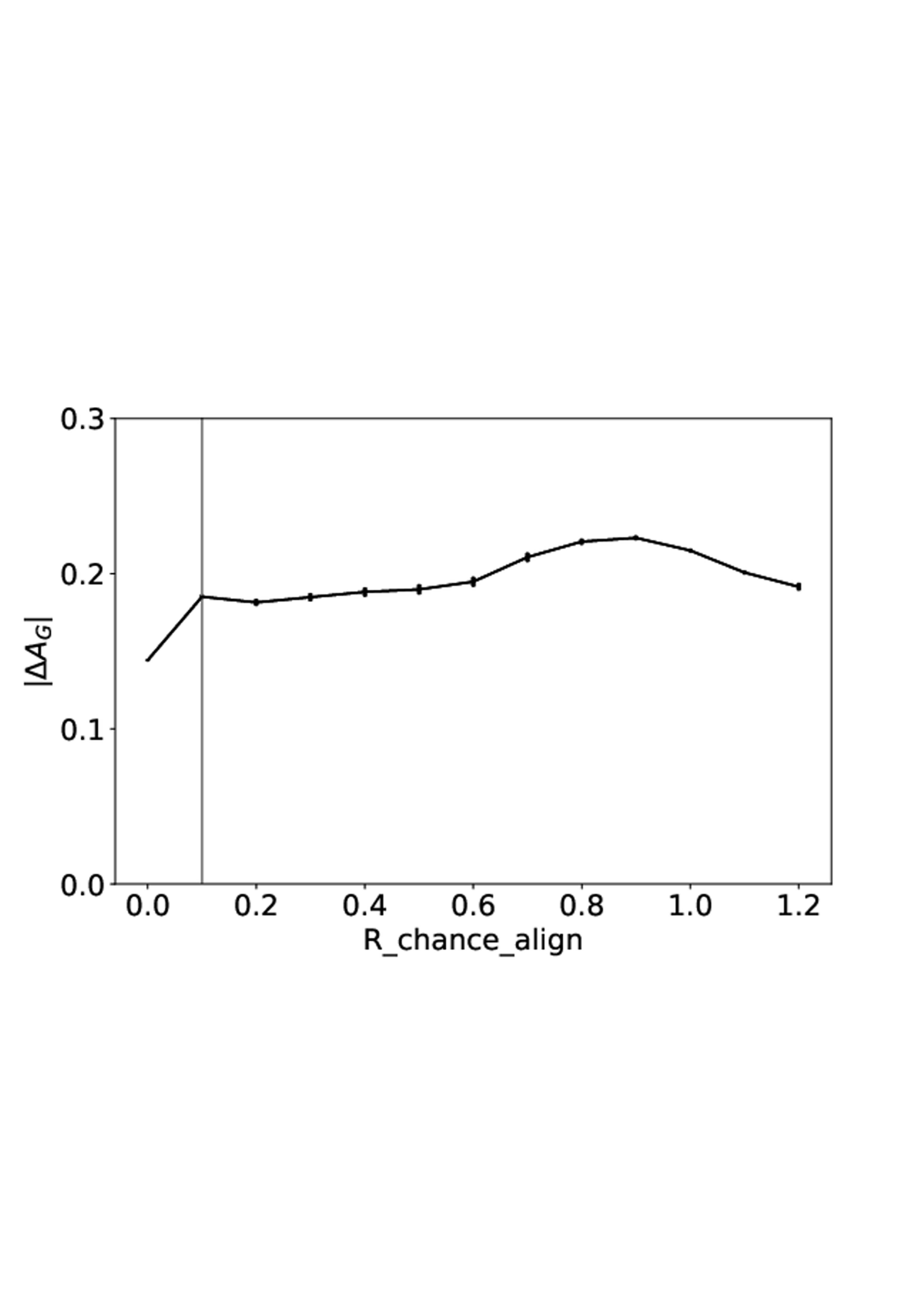}      
  \caption{Relationship between the median modulus of the difference in the values of characteristics independently determined for
the components of a pair and the metric of the probability of a random coincidence of the kinematic characteristics of the components $R\_chance\_align$. Left: the difference in the reduced radial velocities of the components (top) and the difference in
metallicity (bottom). Right: age difference in fractions of the age of the main component (top) and difference in absorption in the
band G (bottom). The vertical gray line marks the maximum value of $R\_chance\_align$ for reliable pairs. Vertical bars indicate
the magnitude of the statistical error.}
  
     \label{FigR}
     \end{figure}

We study the distribution of the median value of the
difference modules $\Delta RV_c$, $\Delta [Fe/H]$, $\Delta Age/Age$, $\Delta A_G$ 
with changes in $R\_chance\_align$. In Fig.~\ref{FigR}, the range
of values of the metric $R\_chance\_align$ is limited to
1.1, which is sufficient to demonstrate the general
trend. Pairs with $R\_chance\_align<0.1$ can be considered as highly probably true. As a characteristic of
the difference in radial velocities $\Delta RV_c$, the reduced
difference in radial velocities of components 1, 2 is
used, considering the dependence of the velocity projection on the coordinates of the
source (l, b). The spatial velocities of the i-components are defined as

\begin{equation*}
    \begin{bmatrix} 
    U^i\\
    V^i\\
    W^i
    \end{bmatrix} = RV^i \cdot \begin{bmatrix}
    \cos  l^i \cos b^i\\
    \sin l^i \cos b^i\\
    \sin b^i
    \end{bmatrix} +\frac{\kappa \mu_l^i}{\varpi^i} \cdot
    \begin{bmatrix}
    -\sin l^i \\
    \cos l^i \\
    0
    \end{bmatrix} + \frac{\kappa \mu_b^i}{\varpi^i} \cdot
    \begin{bmatrix}
    -\cos l^i \sin b^i \\
    -\sin l^i \sin b^i \\
    \cos b^i \\
    \end{bmatrix}
\end{equation*}

Then the radial velocity of one of the components
(to be specific, the first) is calculated, reduced to the
coordinates of component 2:
\begin{equation*}
     RV^1_c=U^1\cos l^2 \cos b^2 + V^1 \sin l^2 \cos b^2  + W^1 \sin b^2.
\end{equation*}

The resulting reduced radial velocity difference $ \Delta RV_c = |RV^1_c-RV^2|$ is free from the projection effect.

Figure~\ref{FigR} shows data characterizing the similarity of
characteristics independently determined for the components of pairs of the catalog \cite{2021MNRAS.506.2269E} within the framework of Gaia DR3. The dependence of this similarity
on the probability of a non-random pairing turns out
to be an interesting diagnostic tool.
Thus, radial velocities (top left panel of Fig.~\ref{FigR})
demonstrate a clear connection with the metric of the
probability of (non)randomness $R\_chance\_align$.
This metric can take values greater than unity, but we
limit the figures to the value $R\_chance\_align=1.1$,
since the most interesting comparison is of pairs of
stars with $R\_chance\_align<0.1$ (“reliable” binaries)
and $R\_chance\_align \approx 1$ (highly probably random
ones). The median difference in reduced radial velocities for reliable pairs (and for pairs up to
 $R\_chance\_align \approx 0.5$) turns out to be 4–6 times
lower than for random combinations of stars.
Of the astrophysical characteristics, the best agreement is shown by the metallicity [Fe/H] of the components (lower left panel of Fig.~\ref{FigR}): for reliable pairs, the median metallicity difference is two times lower
than for random pairs. For the absorption (lower
right panel), the difference in median differences is
less significant, but noticeable.
On the other hand, age estimates from Gaia DR3
show no correlation with the probability of a non-random pairing. The upper right panel shows the median
relative age difference as a fraction of the age of the
brighter component.
The distributions of the differences in the parameters of the components (except for ages) indicate that
a significant proportion of true pairs is present in the
ensemble up to values $R\_chance\_align<0.6$.

  \begin{figure}
     \centering
         \centering
         \includegraphics[width=8cm]{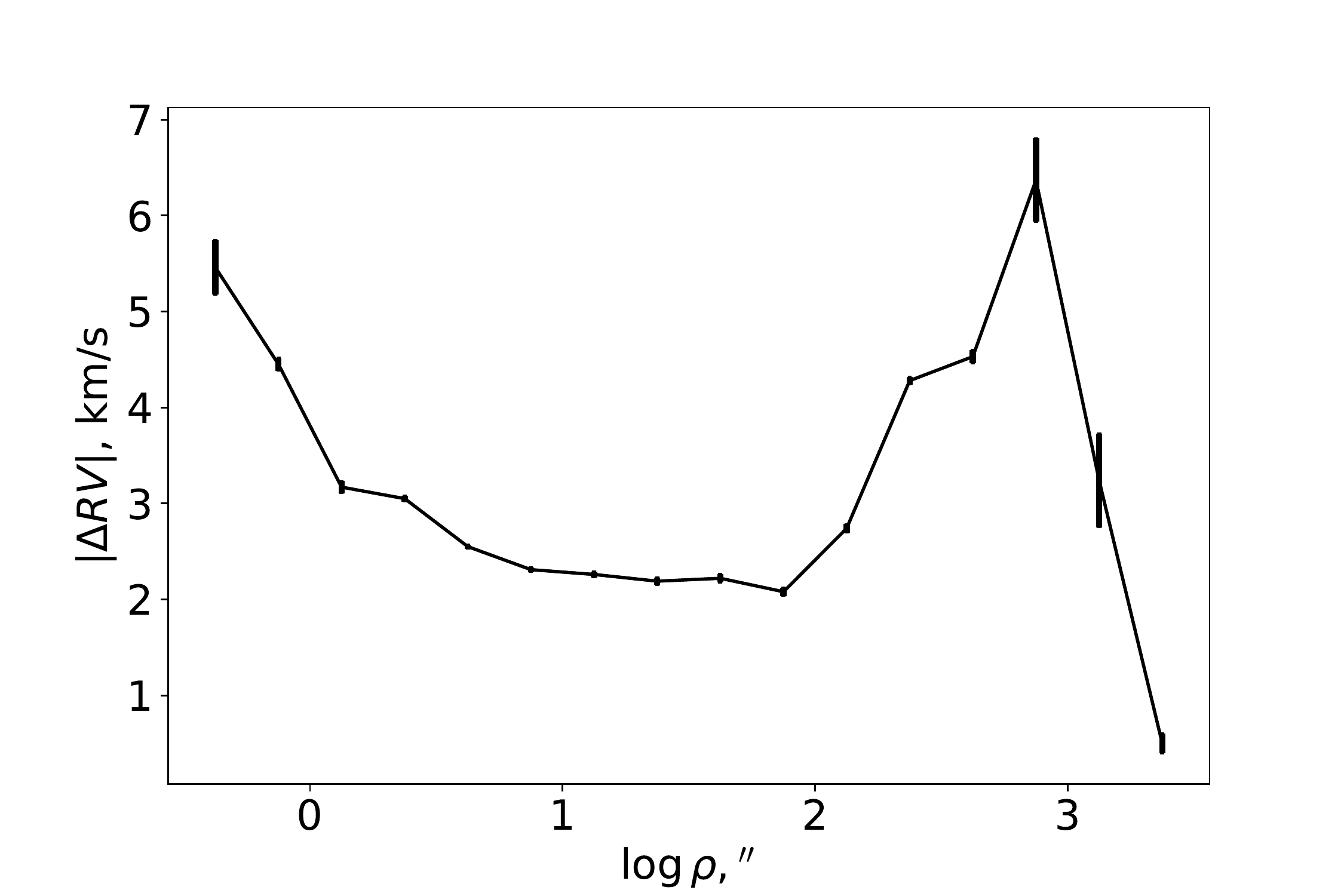}
     \hfill
         \centering
         \includegraphics[width=8cm]{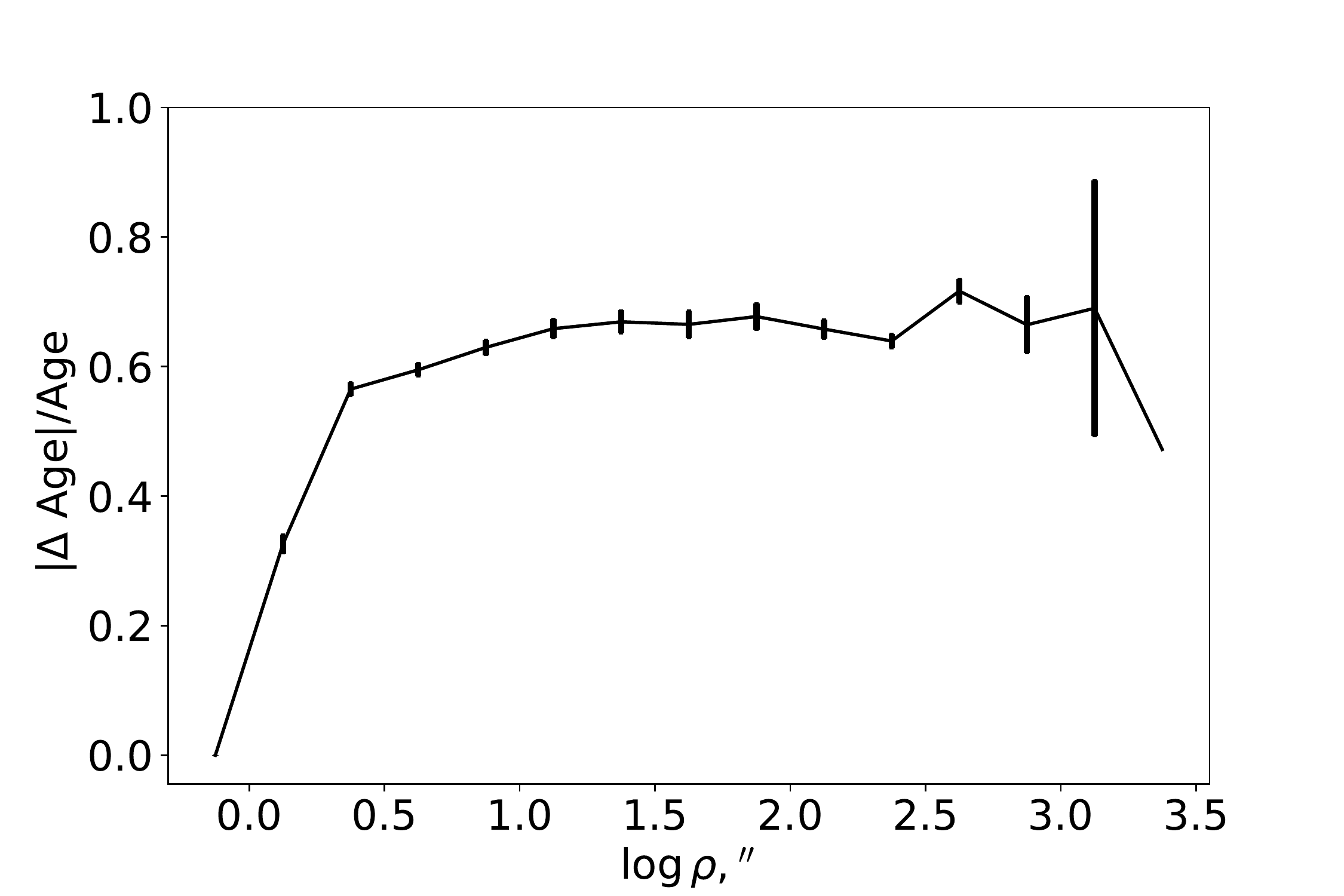}
    \hfill
         \centering
         \includegraphics[width=8cm]{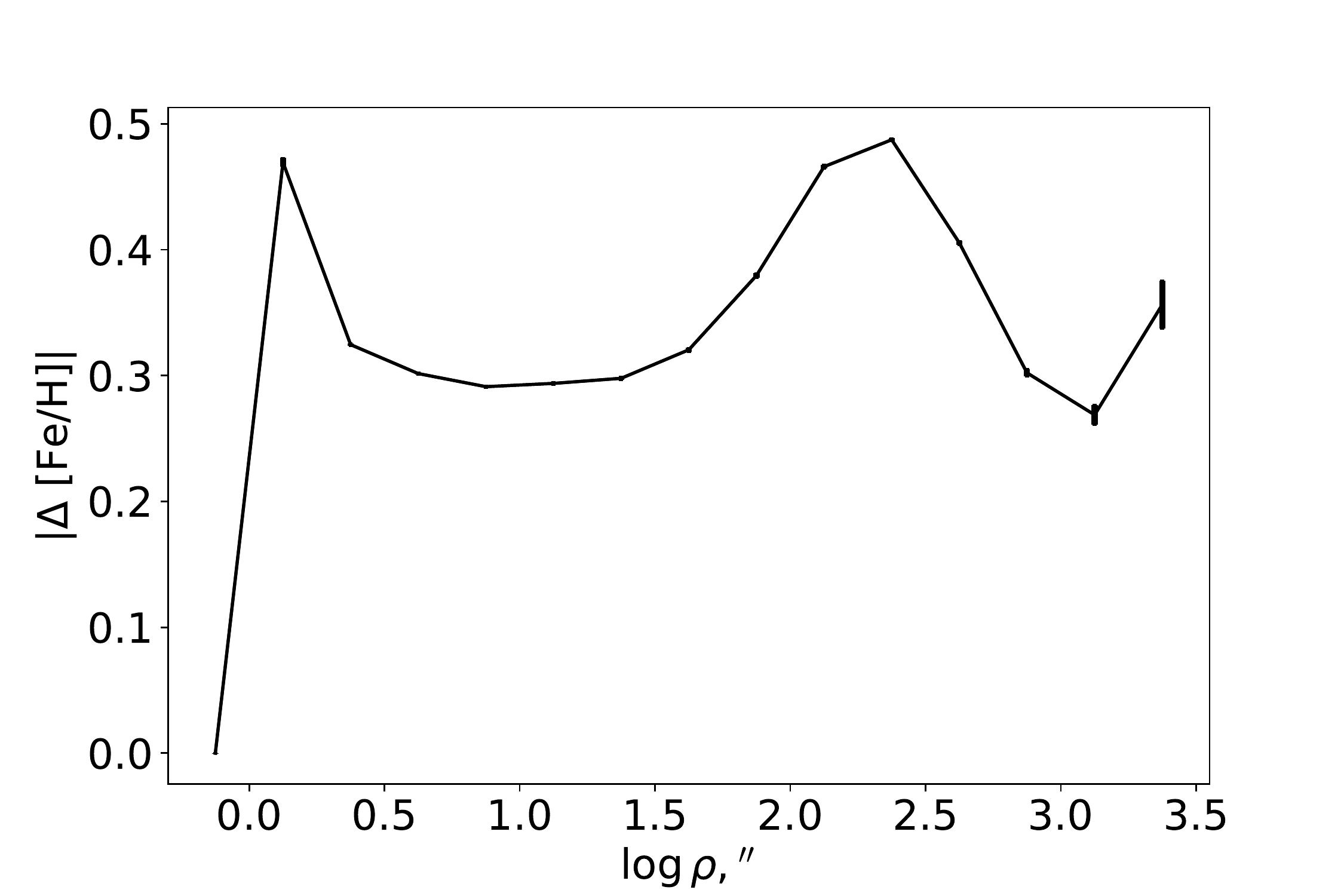}
     \hfill
         \centering
         \includegraphics[width=8cm]{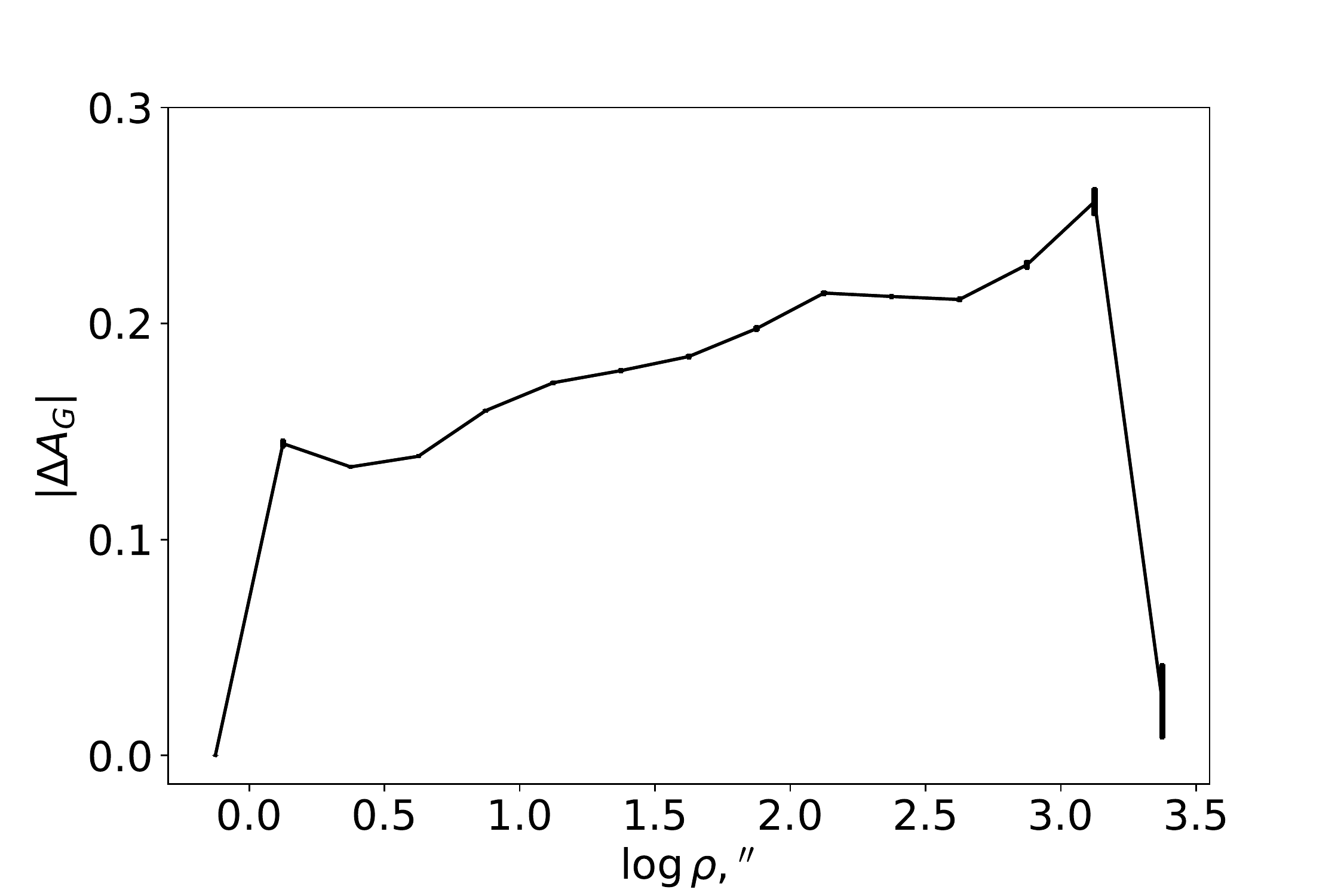}      
  \caption{Relationship between the median modulus of the difference in the values of characteristics independently determined for
the components of a pair and the logarithm of the angular distance between the components (in arcseconds). Left: the difference
in the reduced radial velocities of the components (top) and the difference in metallicity (bottom). Right: age difference in fractions of the age of the main component (top) and difference in absorption in the band G (bottom). Vertical bars indicate the magnitude of the statistical error.}
  
     \label{FigRho}
     \end{figure}
  \begin{figure}
     \centering
         \centering
         \includegraphics[width=8cm]{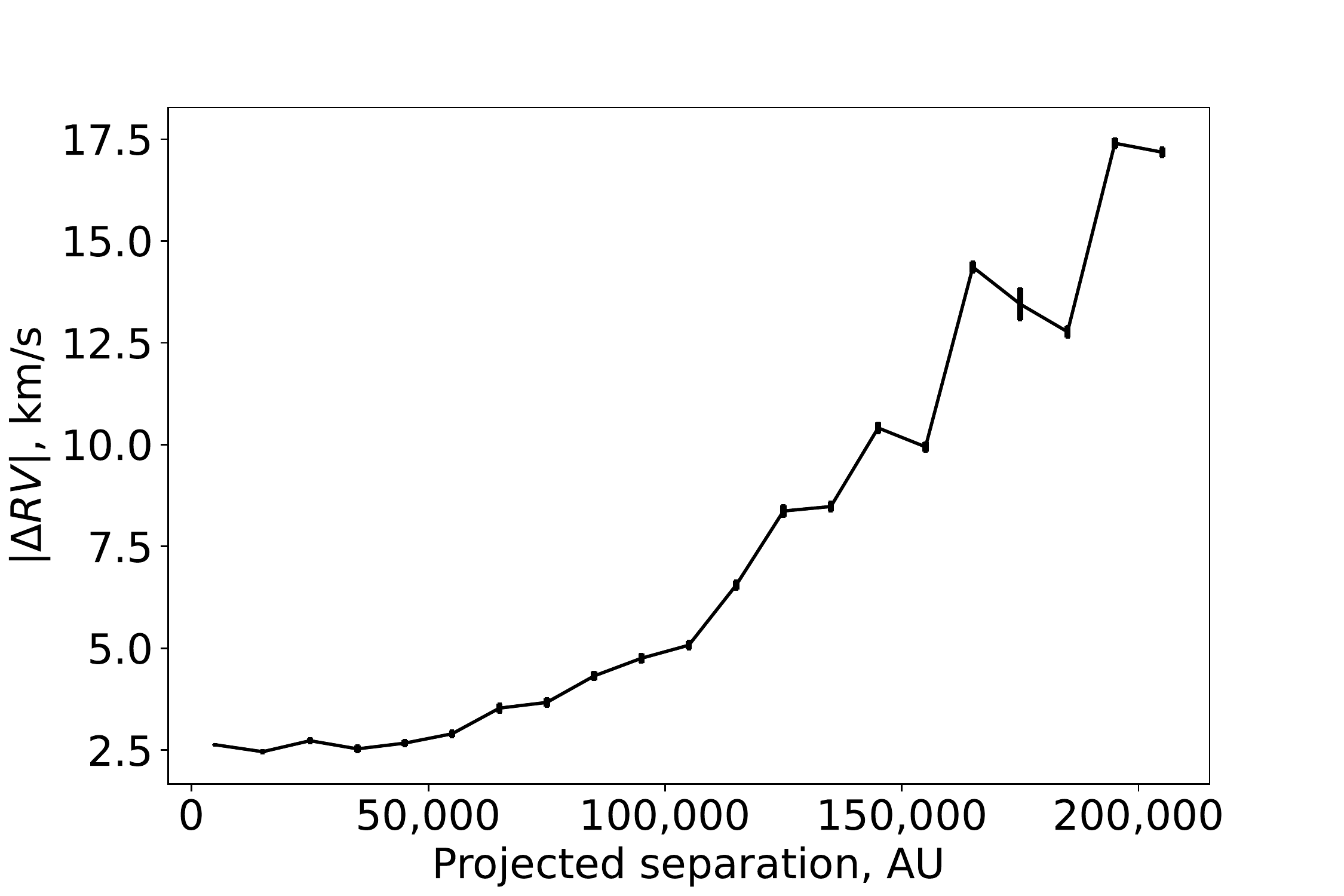}
     \hfill
         \centering
         \includegraphics[width=8cm]{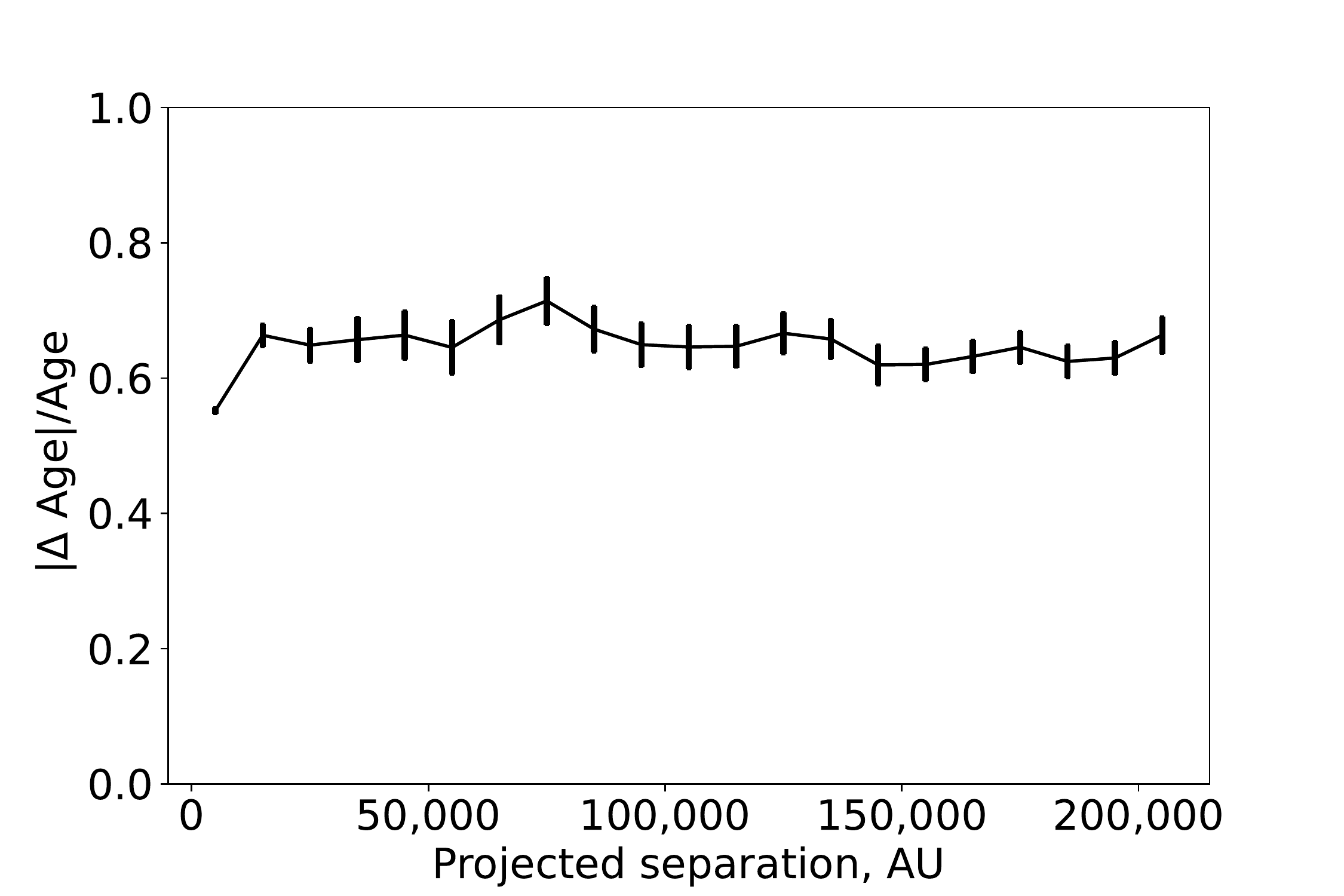}
    \hfill
         \centering
         \includegraphics[width=8cm]{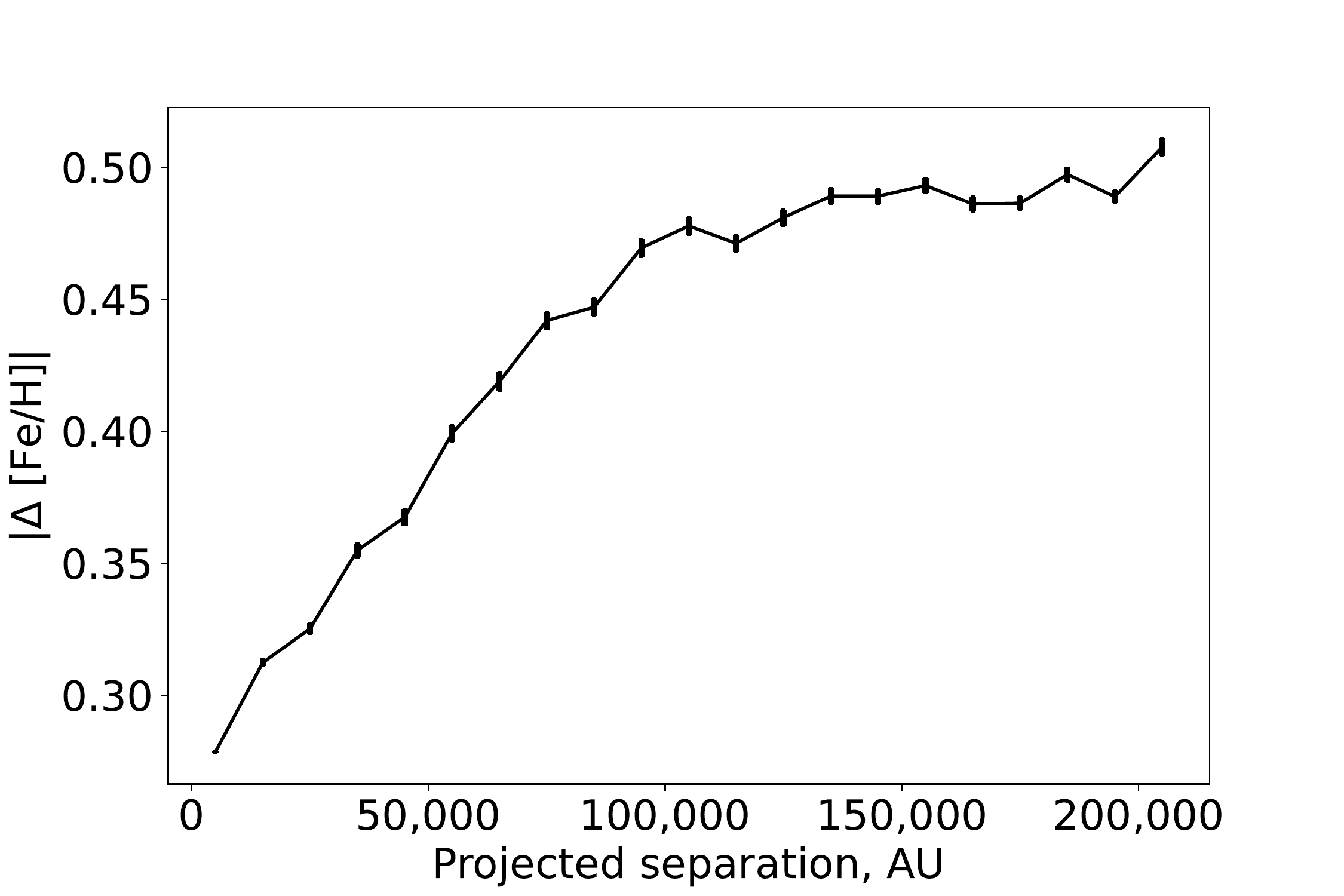}
     \hfill
         \centering
         \includegraphics[width=8cm]{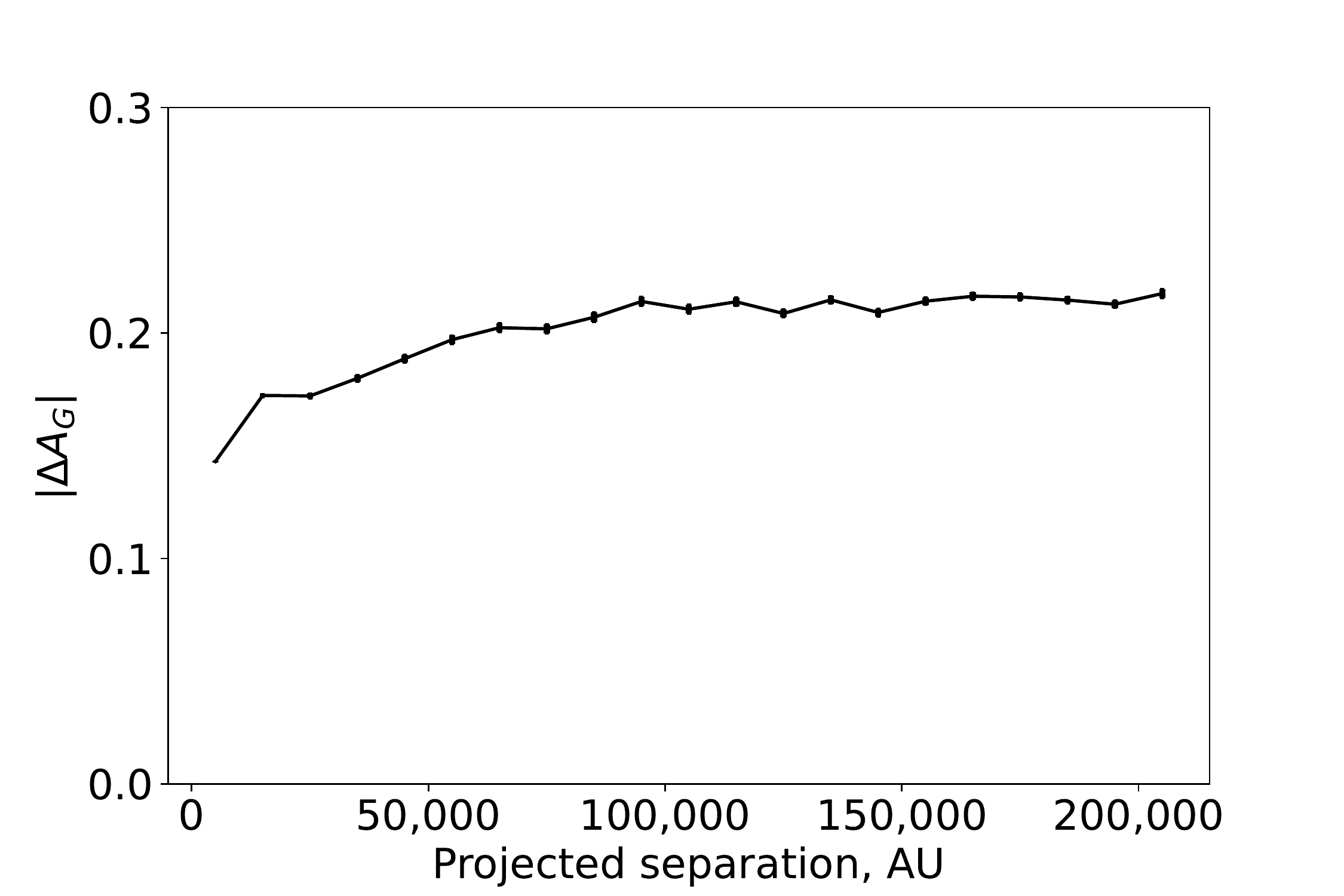}      
  \caption{ Relationship between the median modulus of the difference in the values of characteristics independently determined for
the components of a pair and the projection of the distance between the components onto the tangential plane. Left: the difference in the reduced radial velocities of the components (top) and the difference in metallicity (bottom). Right: age difference in
fractions of the age of the main component (top) and difference in absorption in the G band (bottom). Vertical bars indicate the
magnitude of the statistical error.}
  
     \label{FigS}
     \end{figure}

The quantity $R\_chance\_align$ is statistical in
nature and is related to the observational and physical
characteristics of pairs in the catalog \cite{2021MNRAS.506.2269E} (the angular
distance between the components $\rho$ and the projection
of the physical distance between the components $S_p$)
in non-trivial way. Let us consider the dependence
of $\Delta RV_c$, $\Delta [Fe/H]$, $\Delta Age/Age$, $\Delta A_G$  for the components of the pair in Gaia DR3, on $\rho$ (Fig.~\ref{FigRho}), $S_p$ (Fig.~\ref{FigS}). The angular distance between the components is presented in arcseconds (logarithmic scale), the projection of the linear distance between the components on the tangential plane is in AU (linear scale).
This choice of scales will allow us to further discuss
some interesting effects.

In Fig.~\ref{FigRho}, the increase in value of $\Delta RV_c$ at small values of $\rho$ (upper left panel of the figure) does not correspond to the diagram $\Delta RV_c$ -- $S_p$  (upper left panel of
Fig.~\ref{FigS}) and is apparently associated with a decrease in
the accuracy of determining the radial velocities due to
mutual blending of closely located components. The
identical characteristics [Fe/H], $Age$, $A_G$  при $\rho<1.5''$ of
the components at $\rho<1.5''$ is explained by observational selection: at such angular distances, only
sources with a brightness difference $\Delta G \sim 0$, the so-called “twins”, can be resolved in Gaia. Applying the
same algorithm to two sources with the same observational characteristics leads to identical values of
[Fe/H], $Age$, $A_G$. At $1.5'' \lesssim \rho \lesssim 3''$, however, pairs
with components differing in brightness begin to be
resolved, and the contribution of blending becomes
noticeable, which disappears at high values of $\rho$. It is
interesting that the sharpness of the peak of  $\Delta [Fe/H]$, $\Delta Age/Age$, $\Delta A_G$ associated with blending of components is the more while the dependence of the parameter difference on $R\_chance\_align$ (Fig.~\ref{FigR}) more pronounced: it is significant for $\Delta [Fe/H]$, less pronounced for $\Delta A_G$ and absent for $\Delta Age/Age$. This can be
considered another independent indication of the
comparative internal accuracy of the Gaia DR3 astrophysical parameter estimation method with respect to
different parameters.
As the angular distance between the components
increases, the median values of $\Delta RV_c$, $\Delta [Fe/H]$, $\Delta A_G$
predictably increase (the behavior of the dependencies 
for $\rho \gtrsim 1000''$ is statistically insignificant). At the
same time, the median value of $\Delta Age/Age$ remains
practically unchanged.

In Fig.~\ref{FigS}, the behavior of the dependence of
median values of $\Delta RV_c$, $\Delta [Fe/H]$, $\Delta Age/Age$, $\Delta A_G$  on
the projection of the physical distance between the
components shows a picture qualitatively similar to
Fig.~\ref{FigR}. The median difference in values increases significantly with increasing distance between the components for radial velocities and metallicities (left
upper and lower panels, respectively), indicating an
increase in the fraction of non-physical pairs; at the
same time, the median difference $\Delta A_G$ (lower right
panel) grows slightly, while is approximately constant.
Thus, the distributions of the differences in the
astrophysical parameters of the components depending on $R\_{chance}\_{align}$, $\rho$ and $S_p$, indicate better
agreement between the characteristics of the components for metallicities, noticeable but weak agreement
for $A_G$, and almost no agreement for ages.
The low general accuracy of component age estimates can be attributed to the fact that for main
sequence stars, the observational characteristics, considering their errors, depend very little on age. Moreover, for a subsample that includes 400 reliable pairs
with components in which both components on the
Hertzsprung–Russell diagram look evolved, the
median value is $\Delta Age/Age =0.25$, while the median
value for the entire ensemble of reliable pairs is
$\Delta Age/Age =0.61$. It can therefore be expected that for
stars leaving the main sequence and for giant stars, age
estimates in Gaia DR3 are more reliable.

\subsection{Study of the Spatial Completeness of the Catalog}

\begin{figure}
     \centering
         \centering
         \includegraphics[width=8cm]{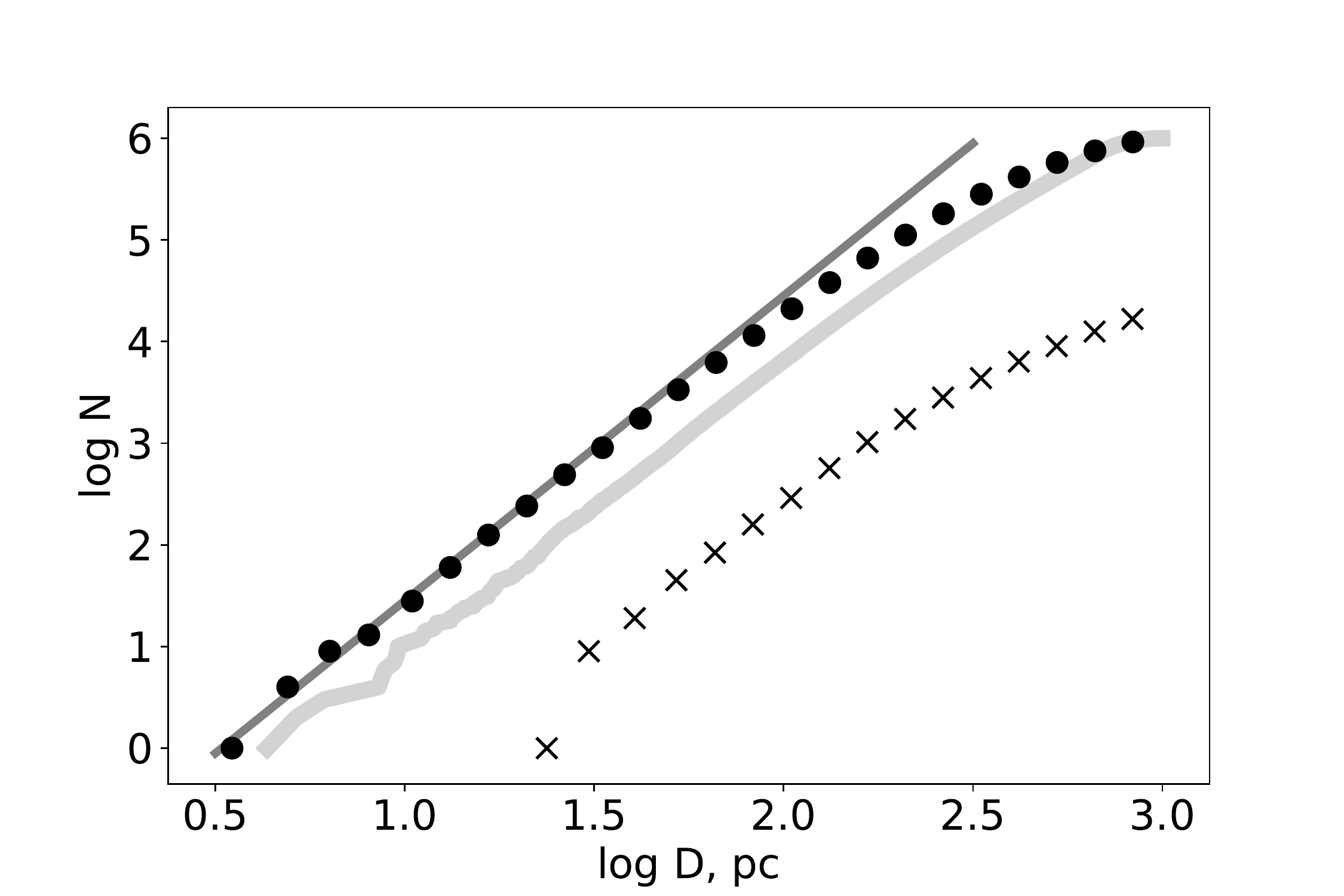}
     \hfill
         \centering
         \includegraphics[width=8cm]{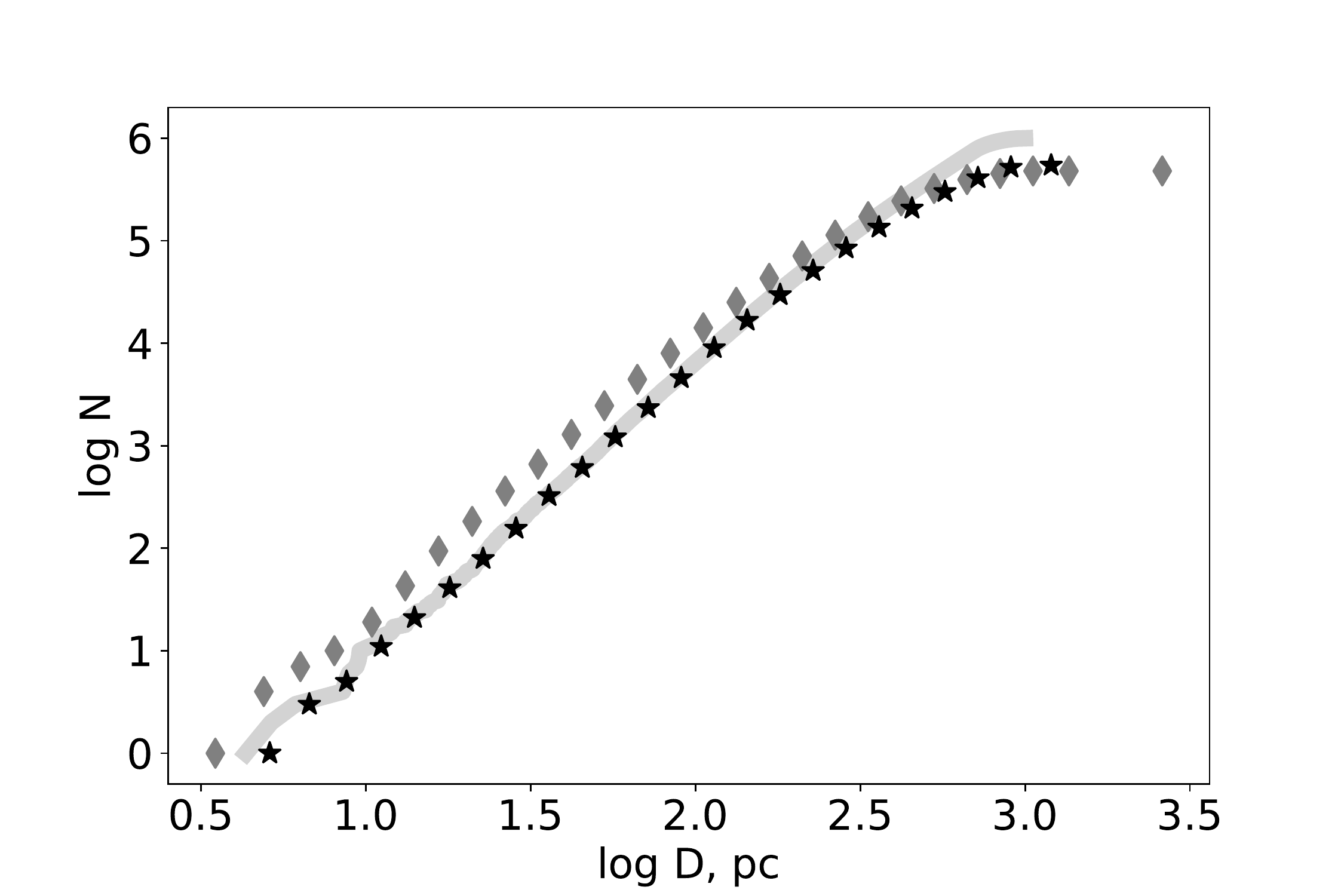}
  
  \caption{Number of stars (main components of pairs) as a function of distance from the Sun. Left: black circles are all reliable pairs,
black oblique crosses are pairs with an absolutely bright main component ($G_{abs}<2^m$). The light gray line is the model distribution
(see text), the dark gray line is the function $f(x) \propto 3x$, for comparison. Right: black stars are pairs with absolute magnitude of the
main component $G_{abs}<7^m$, gray diamonds are pairs with absolute magnitude of the main component $G_{abs}\geq 7^m$, light gray line
is the model distribution.}
      \label{FigSpaceCompl}
     \end{figure}

\begin{figure}
     \centering
         \centering
         \includegraphics[width=8cm]{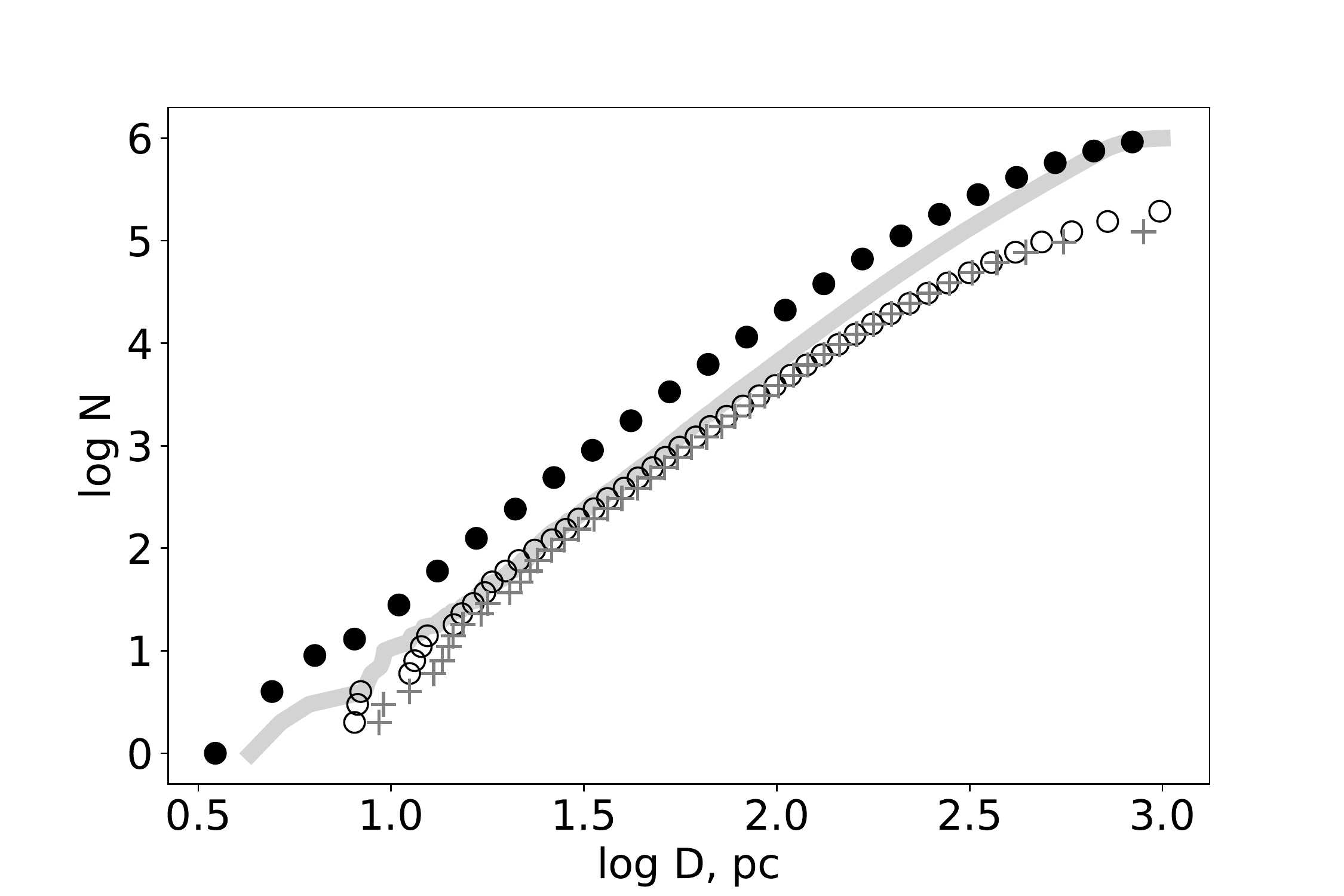}
     \hfill
         \centering
         \includegraphics[width=8cm]{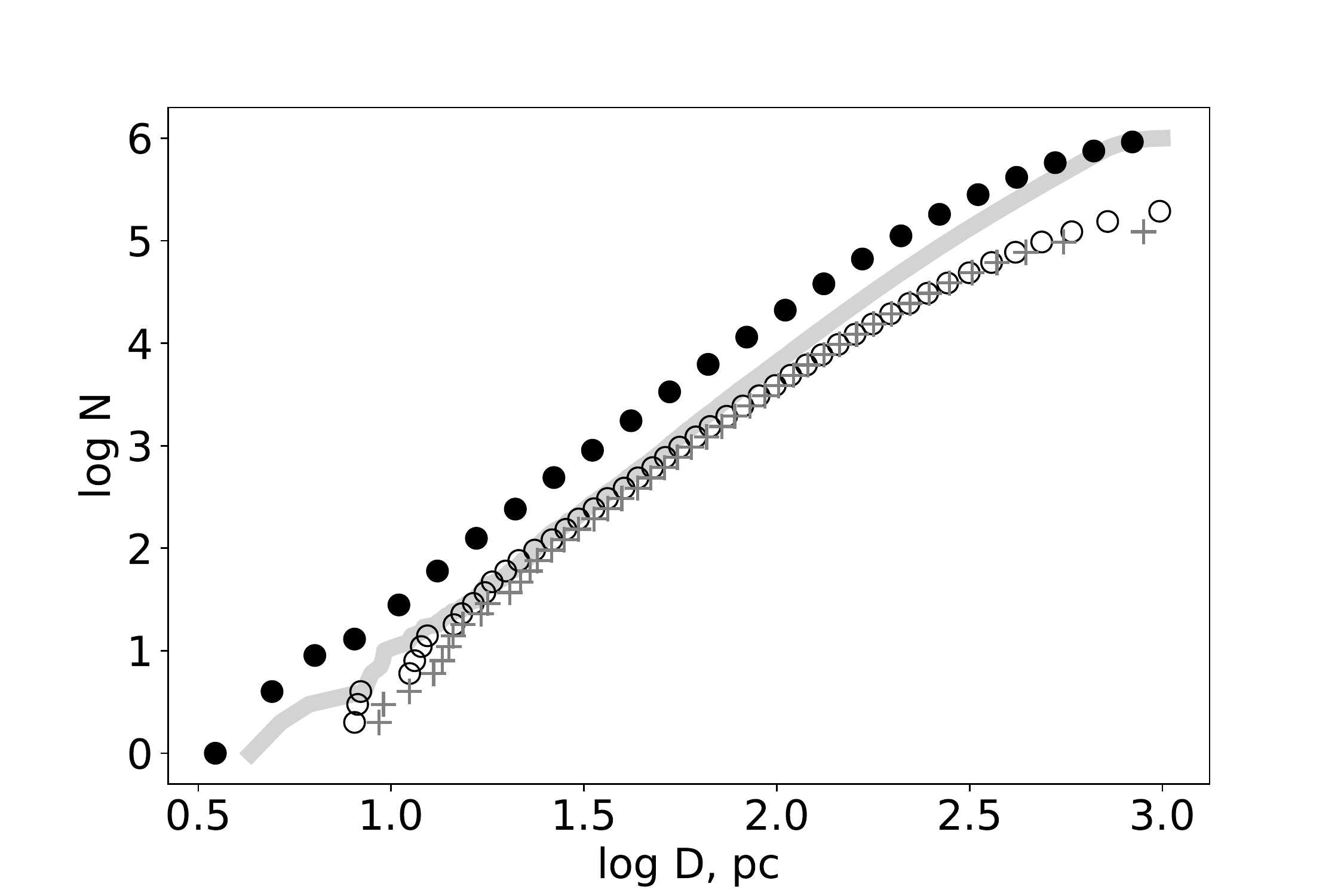}
  
  \caption{Number of stars (main components of pairs) as a function of distance from the Sun. Left: black circles are all reliable pairs,
gray crosses are pairs with a difference in the brightness of the components $\Delta G>5^m$, black empty circles are pairs with a difference in the brightness of the components $\Delta G<0.5^m$. Light gray line is the model distribution (see text). Right: all pairs and
model distribution as in the left panel. Gray squares are pairs with a distance between components (projection) $a_{proj}>1000$ AU, black empty triangles – with a distance between components $a_{proj}<100$ а.е.}
     \label{FigParamCompl}
     \end{figure}

Based on the principles of creating the catalog \cite{2021MNRAS.506.2269E} (excluding regions of open star clusters from consider and discusstion of the Fig.~\ref{FigAngDist} in the Section~\ref{data} 
one can
expect that the catalog of wide binaries does not have
spatial completeness in the volume of 1 kpc. Let us
investigate the connection between its spatial incompleteness and the characteristics of binary stars by
modifying the method described in \cite{2016A&A...585A.101K} for studying
the dependence of the spatial completeness of an
ensemble of star clusters on their integral stellar magnitude.
Provided that the space around the Sun is uniformly filled with binary stars, the asymptote of the
distribution describing the increase in their number
with distance to the pair (on a logarithmic scale)
should be a straight line $f(x) \propto 3x$. However, the catalog under study contains stars at a distance of up to
1~kpc, which is noticeably greater than the height scale
of the stellar disk. We model the expected distribution
of stars, distributing them uniformly in galactic rectangular coordinates (X,Y), and along the coordinate Z
—-- in accordance with the law with a height scale
of 150 pc, characteristic of G-dwarfs \citep{2017MNRAS.470.1360B}. The normalization of the distribution is such that there are
1 million stars at a distance of 1 kpc from the Sun,
which approximately coincides with the number of
reliable pairs in the catalog. In Figs.~\ref{FigSpaceCompl}, \ref{FigParamCompl}, the model distribution is shown by a pale gray wide line. The distribution of all reliable ($R\_chance\_align<0.1$) pairs
in the catalog is shown in large black circles (left panel
of Fig.~\ref{FigSpaceCompl}, both panels of Fig.~\ref{FigParamCompl}). In all cases, the probable geometric distance to the main component is
used to estimate the distance to the pair \citep{2021AJ....161..147B}.

Figure~\ref{FigSpaceCompl} shows the dependence of the spatial completeness of the catalog on the absolute magnitude of
the main component. The left panel shows the distribution of all stars in comparison with the model distribution and, for the support, the asymptote $f(x) \propto 3x$
(thin gray line). The oblique crosses show the change
with distance in the number of stars of absolute
magnitude $G_{abs}<2^m$. This distribution is expectedly
incomplete near the Sun (due to the Gaia limitations
for bright magnitudes). Comparison of the complete
distribution of reliable catalog pairs with the model
one (the difference is about 0.4 dex near the Sun)
allows us to estimate the probable shortage of stars in
the catalog. If it were spatially complete, the expected
number of pairs in a volume of 1 kpc would be about
2.5 million. The catalog becomes incomplete in the
region $D>200$~pc from the Sun, which coincides with
the authors’ own estimate, made by a different
method.
The right panel compares distributions for stars
with absolute magnitude $G_{abs}<7^m$ and $G_{abs}>7^m$. For
main sequence stars corresponds to spectral class $K6$, according to \url{https://www.pas.rochester.edu/~emamajek/EEM_dwarf_UBVIJHK_colors_Teff.txt} \citep{2013ApJS..208....9P}. 
The more absolutely bright part of the
ensemble (black dots) maintains a constant spatial
density up to 400 pc, while the spatial density of the
more absolutely faint pairs begins to decrease, starting
from about 50 pc from the Sun.
Figure~\ref{FigParamCompl} demonstrates the influence of pair characteristics on the change in spatial density of pairs with
distance. The left panel demonstrates that there is
practically no dependence on the difference in the
magnitudes of the components: the change in the
number of stars with high contrast (gray crosses) and
stars with almost identical components (black dots)
with distance is almost identical. In the immediate
vicinity of the Sun, a small deficit of pairs with a
brightness difference between the components $\Delta G>5^m$
may be due to the incomplete representation
of bright stars in the Gaia catalog, and at distances of
about 1 kpc —-- to the fact that the secondary components of such pairs become too faint.
In the right panel of Fig.~\ref{FigParamCompl}, gray and black dots
show the relation of the change in spatial density of
pairs with distance from the Sun depending on the distance between the components. In all cases, the projection of distance onto the tangential plane is used as
the distance. Gray dots indicate the spatial distribution of pairs with $a_{proj}>1000$~AU, and black dots indicate pairs with the distance between the components $a_{proj}<100$~AU. The relative lack of “close” binaries
with  $a_{proj}<100$~AU starts from about 25 pc from the
Sun due to the fact that some pairs with high contrast
cease to be resolved in the Gaia catalog, and beyond
125 pc there are almost no such pairs. As the distance
from the Sun increases, the widest pairs begin to dominate in the ensemble.

\section{DISCUSSION}
\label{disc}

  \begin{figure}
     \centering
         \centering
         \includegraphics[width=8cm]{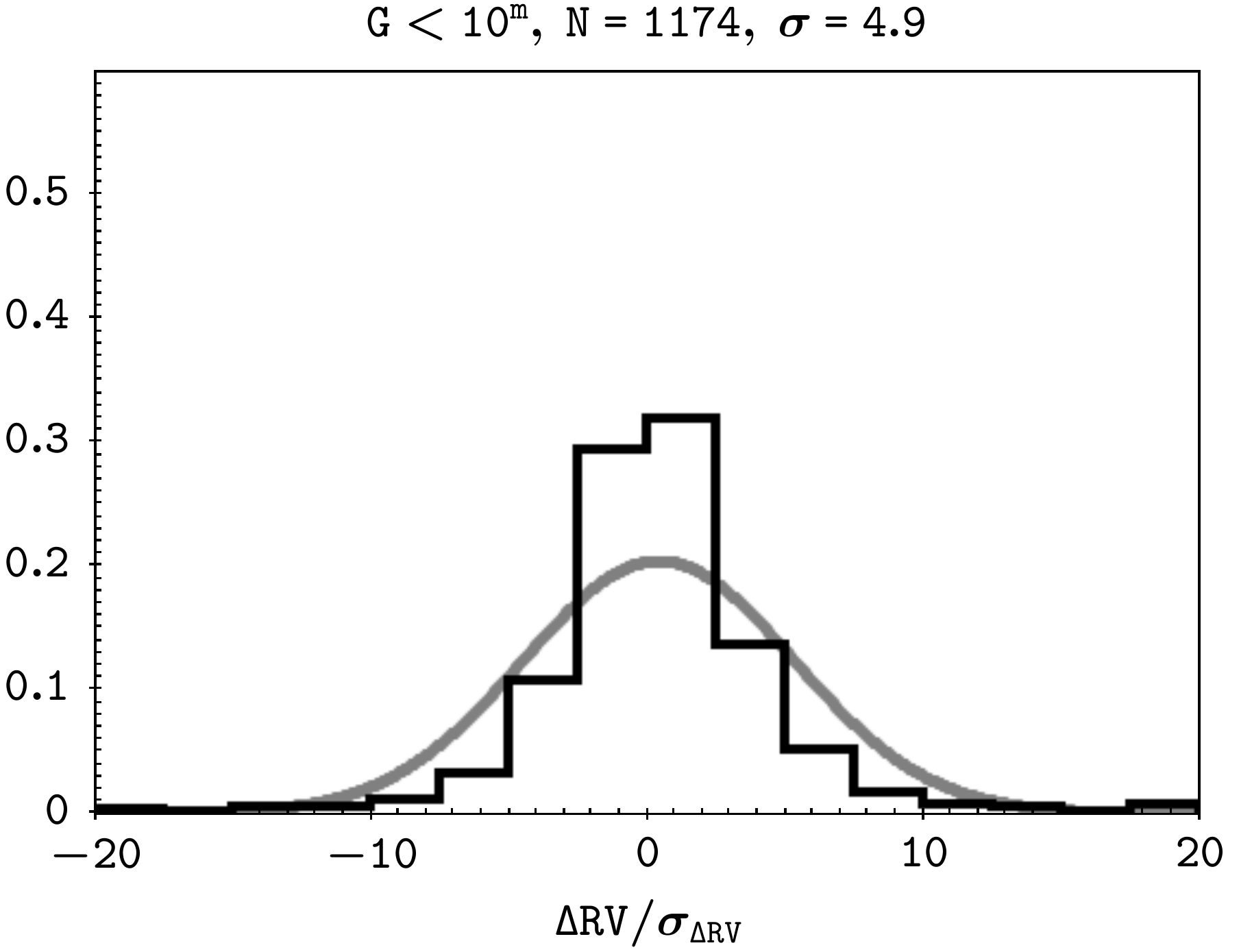}
     \hfill
         \centering
         \includegraphics[width=8cm]{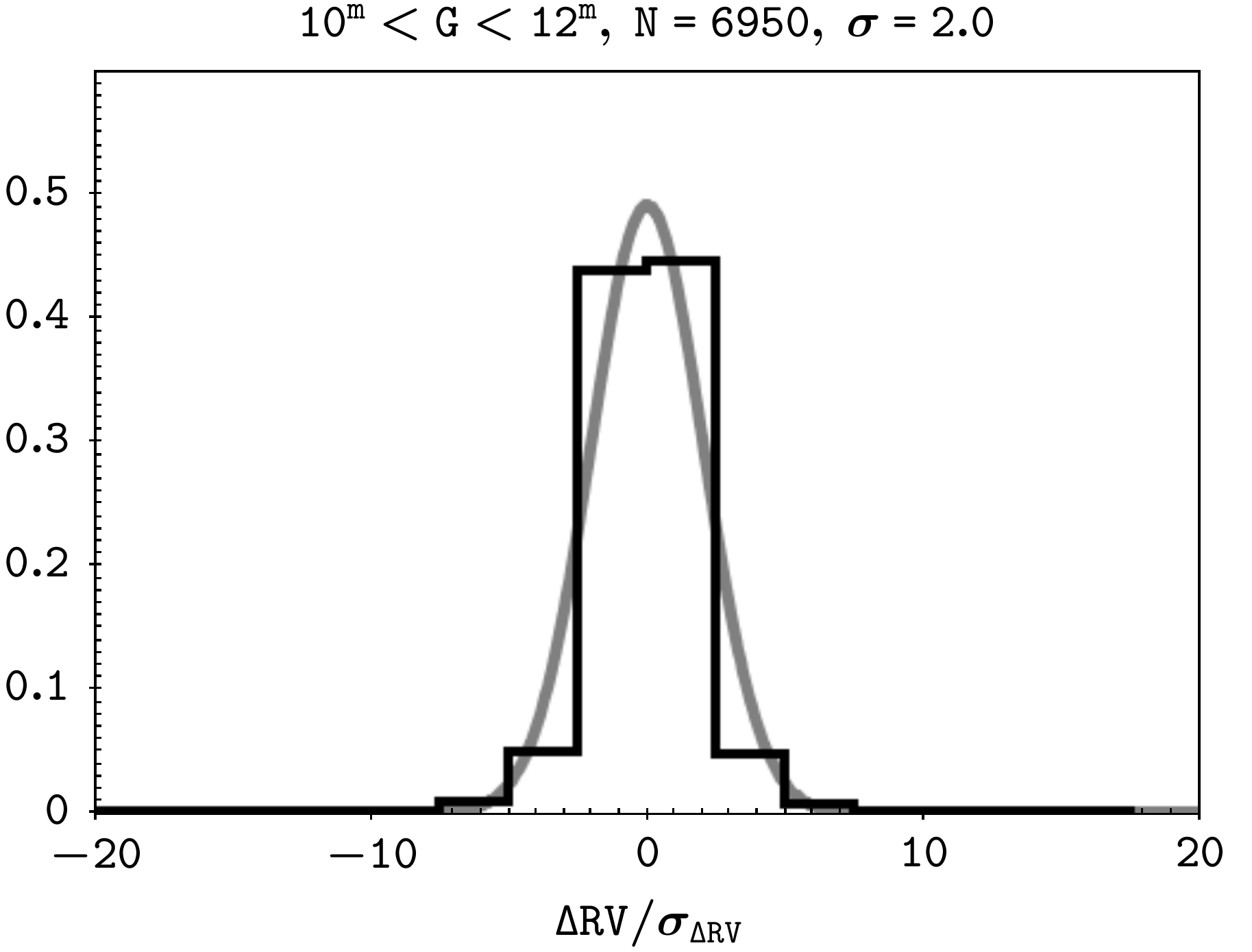}
    \hfill
         \centering
         \includegraphics[width=8cm]{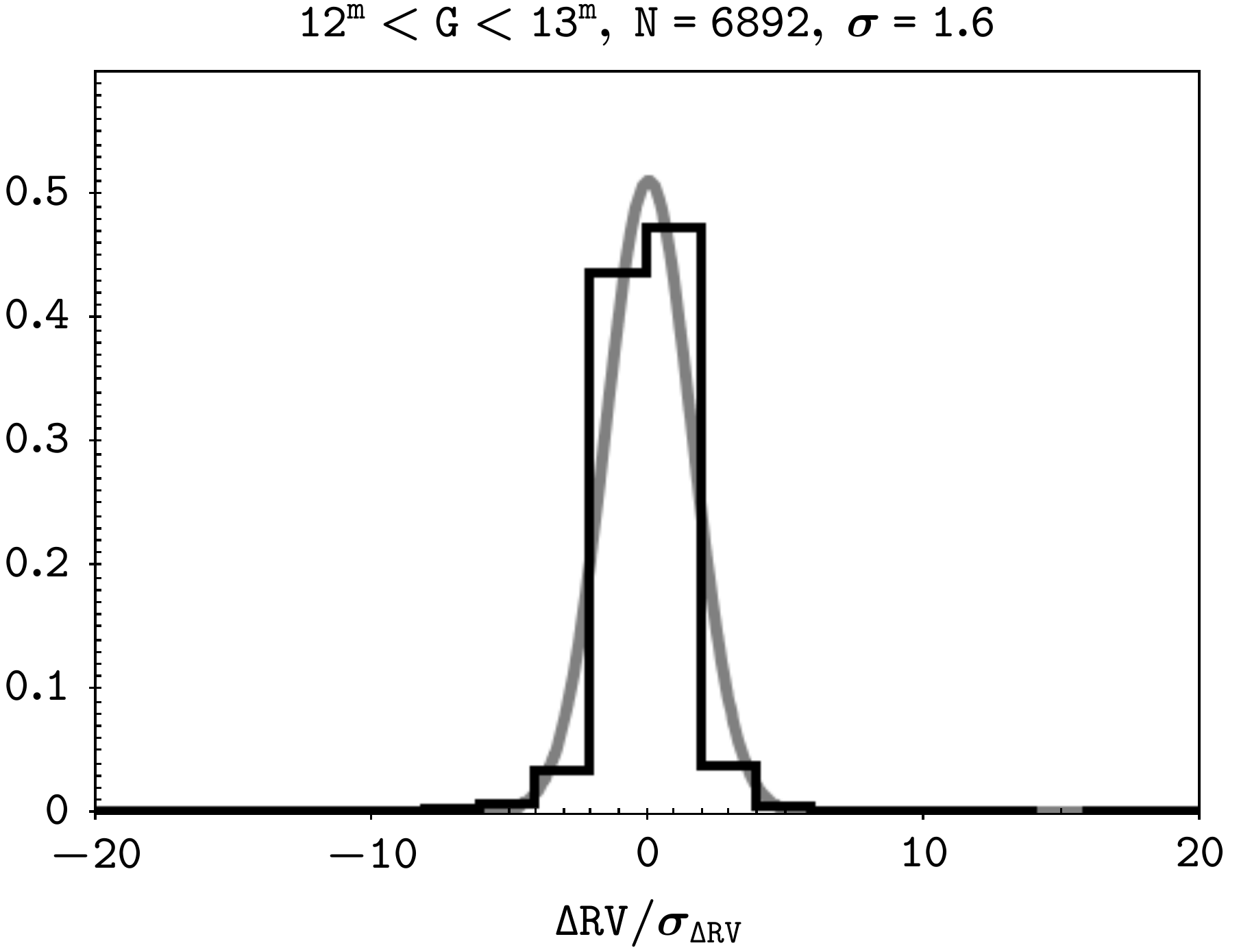}
     \hfill
         \centering
         \includegraphics[width=8cm]{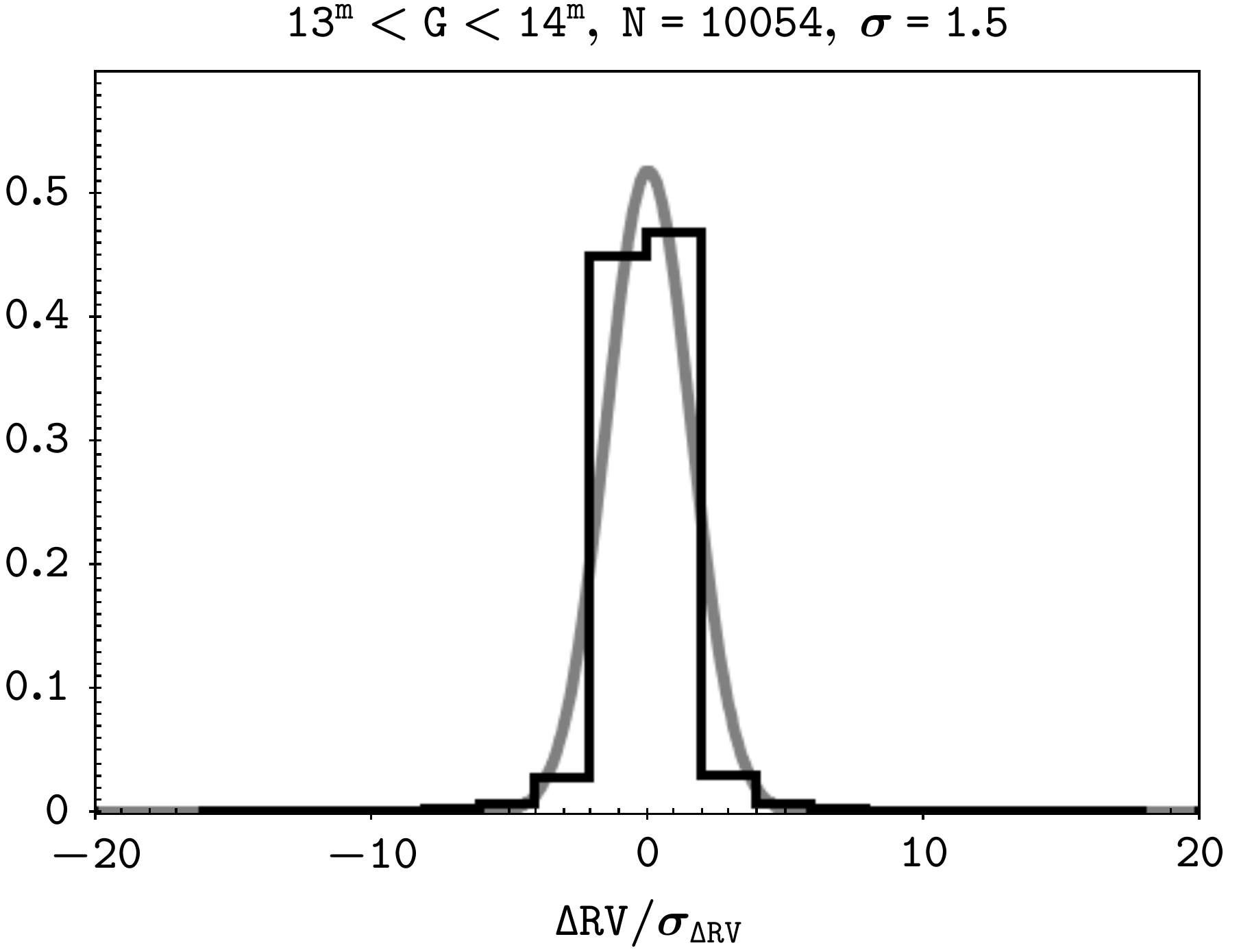}   
             \hfill
         \centering
         \includegraphics[width=8cm]{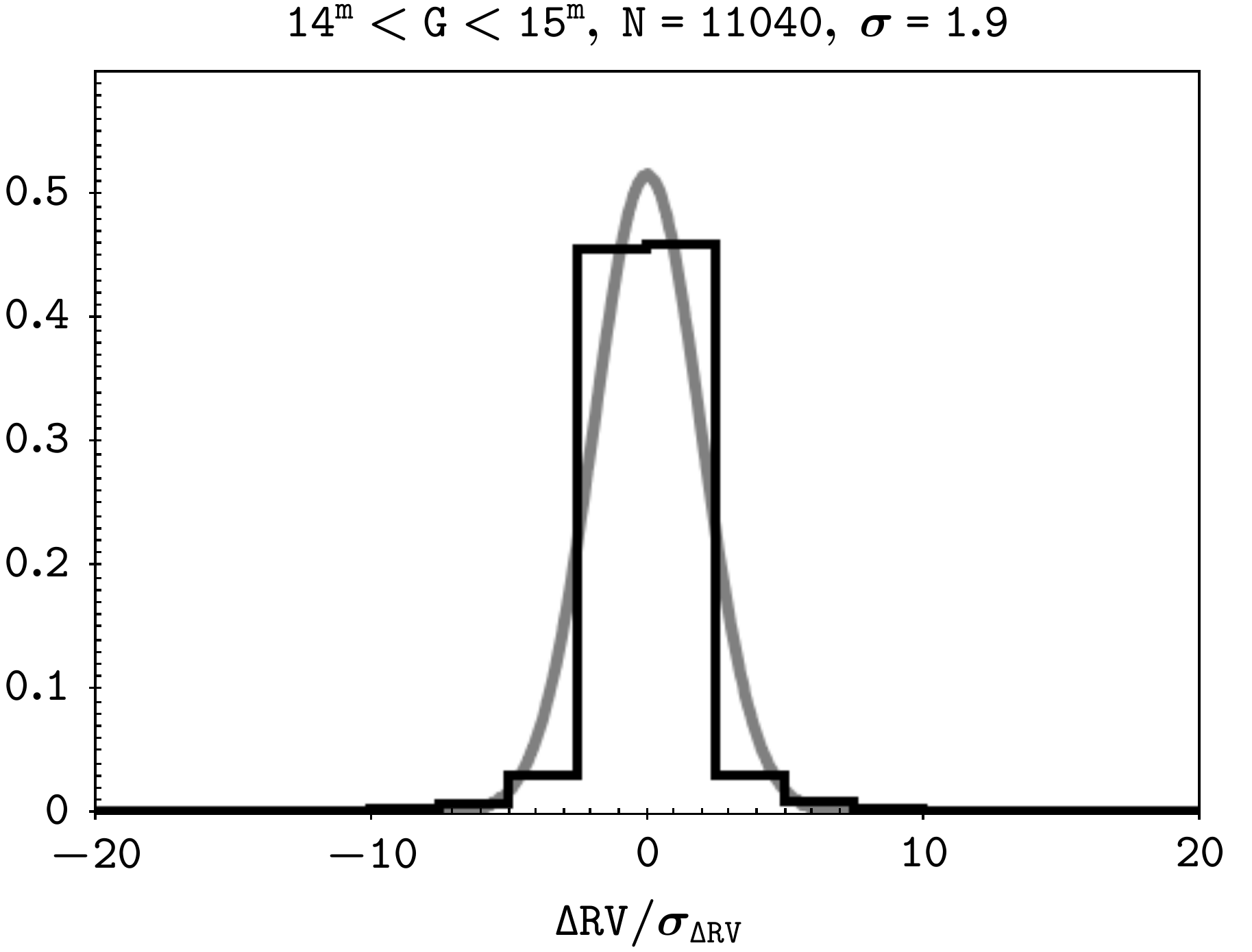}
     \hfill
         \centering
         \includegraphics[width=8cm]{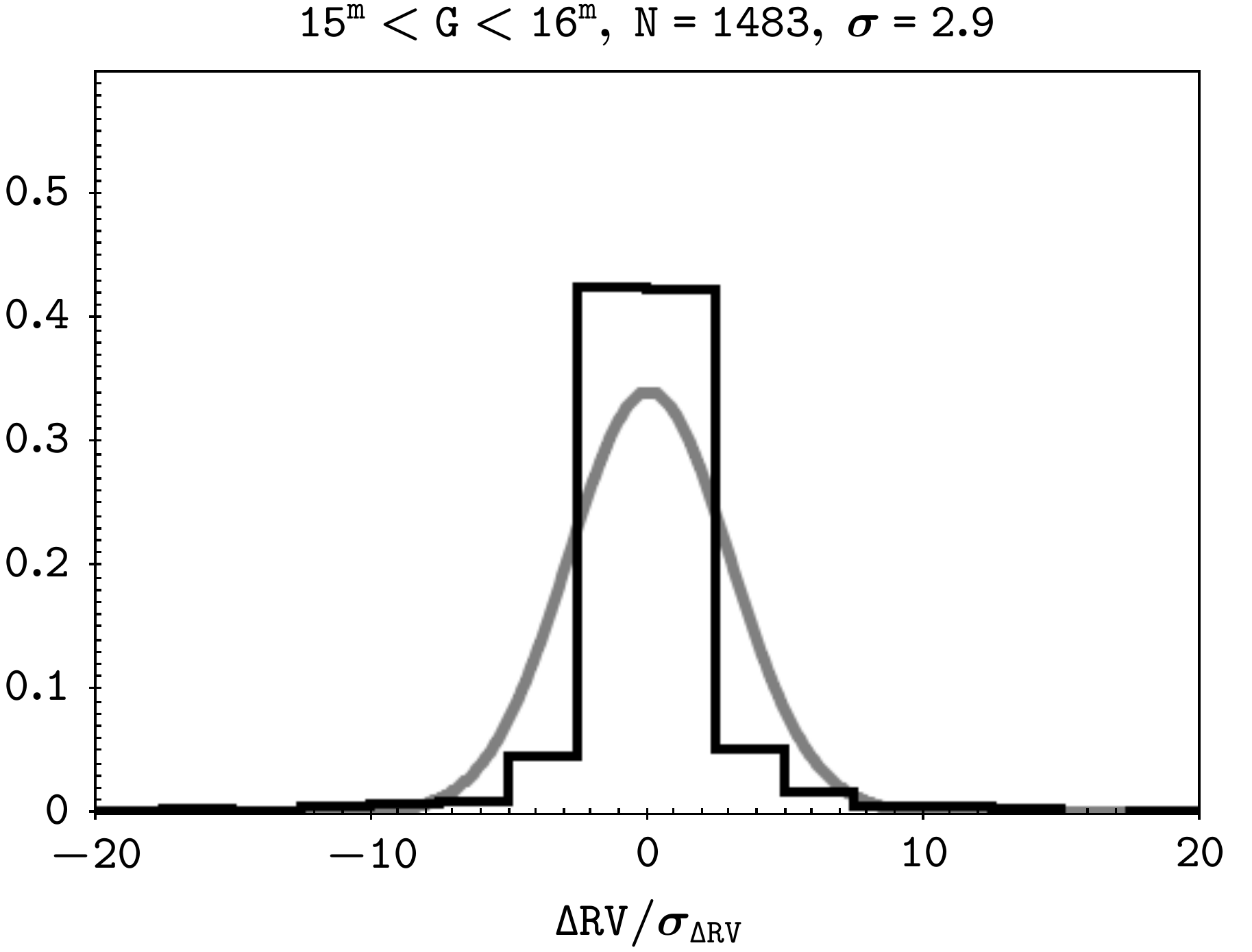}   
  \caption{Distribution of the difference in radial velocities of the components of the pair, normalized to error, for stars in different
ranges of visible magnitude G. A pair is included in the range statistics if both components have a magnitude within the specified
range. Black histograms are observed distributions, gray lines are their approximation by Gaussians. Above each panel is the range
of stellar magnitudes, the number of pairs participating in the statistics, and the standard deviation characterizing the Gaussian
approximation of the distribution. }
  
     \label{FigdRV}
     \end{figure}

\begin{table}[b]
 \begin{center}
 \caption{Median radial velocity errors depending on the magnitude, in km/s.}
 \label{tableRV}
 \medskip
\begin{tabular}[c]{|c|c|c|c|c|}
\hline
G, mag& $\sigma_{RV}\_{DR3}$ & $\sigma_{RV}\_{bin}$ & $\sigma_{RV}\_{orb}$ & $\sigma_{RV}\_{exp} $    \\

\hline

<10    & 0.33 & 1.6     & 1.2 & 1.1\\
10-12    & 0.86 & 1.7     & 0.90 & 1.5\\
12-13     & 1.9 & 3.0     & 0.81 & 2.9\\
13-14     & 3.2 & 4.8     & 0.76 & 4.7\\
14-15     & 5.5 & 10.5    & 0.70 & 10.5\\
15-16     & 6.7 & 20.1     & 0.67 & 20.1\\

\hline 
\end{tabular}
\end{center}
The median errors are given for different ranges of magnitude :
the radial velocity from Gaia DR3 ($\sigma_{RV}\_{DR3}$), the standard deviation of the normalized radial velocity difference between the
components of a pair ($\sigma_{RV}\_{bin}$), the median estimate of the possible radial velocity difference of a pair associated with orbital
motion in a binary system ($\sigma_{RV}\_{orb}$), and our resulting estimate
of the expected characteristic radial velocity error ($\sigma_{RV}\_{exp} $).
\end{table}

  \begin{figure}
     \centering
         \centering
         \includegraphics[width=7cm]{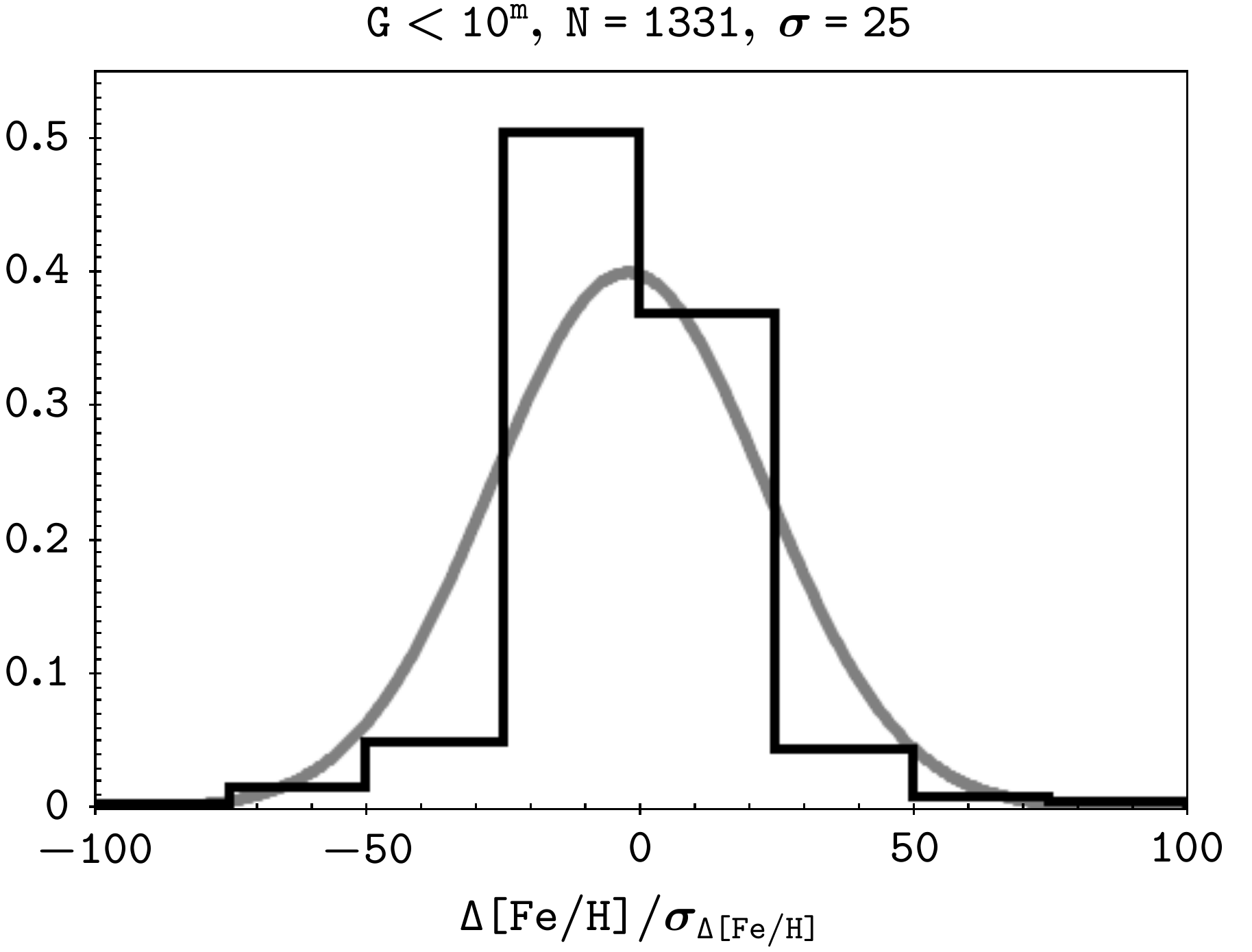}
     \hfill
         \centering
         \includegraphics[width=7cm]{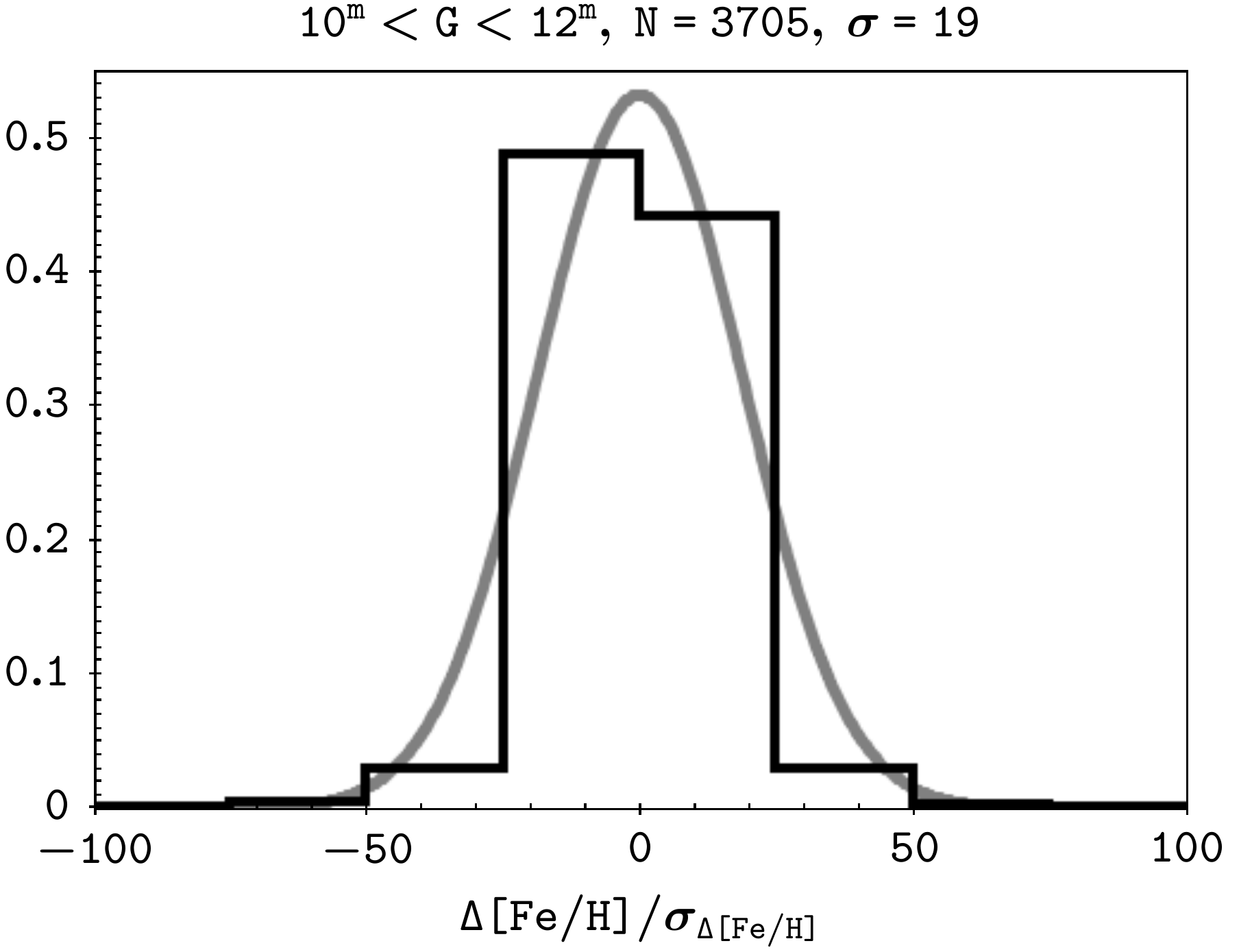}
    \hfill
         \centering
         \includegraphics[width=7cm]{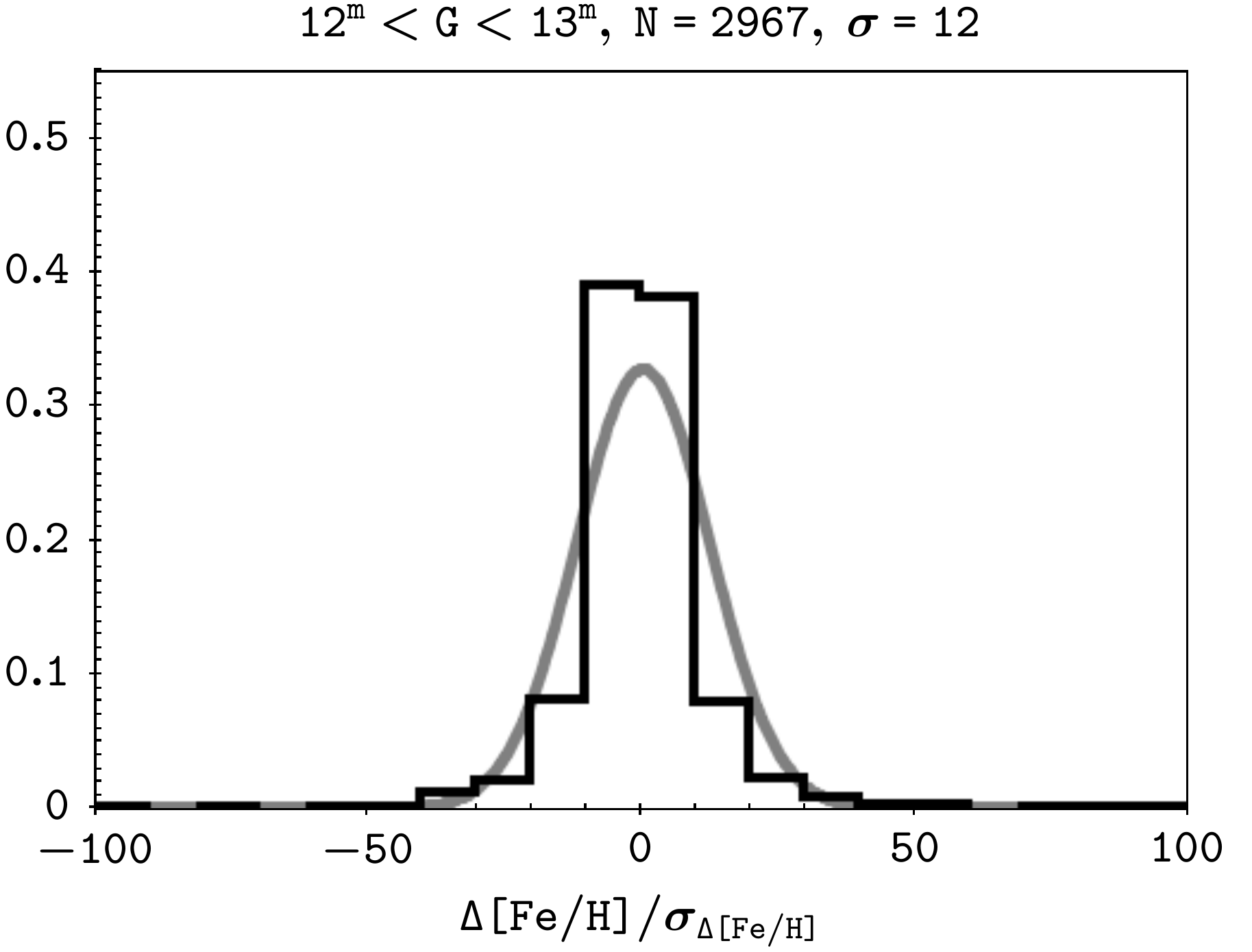}
     \hfill
         \centering
         \includegraphics[width=7cm]{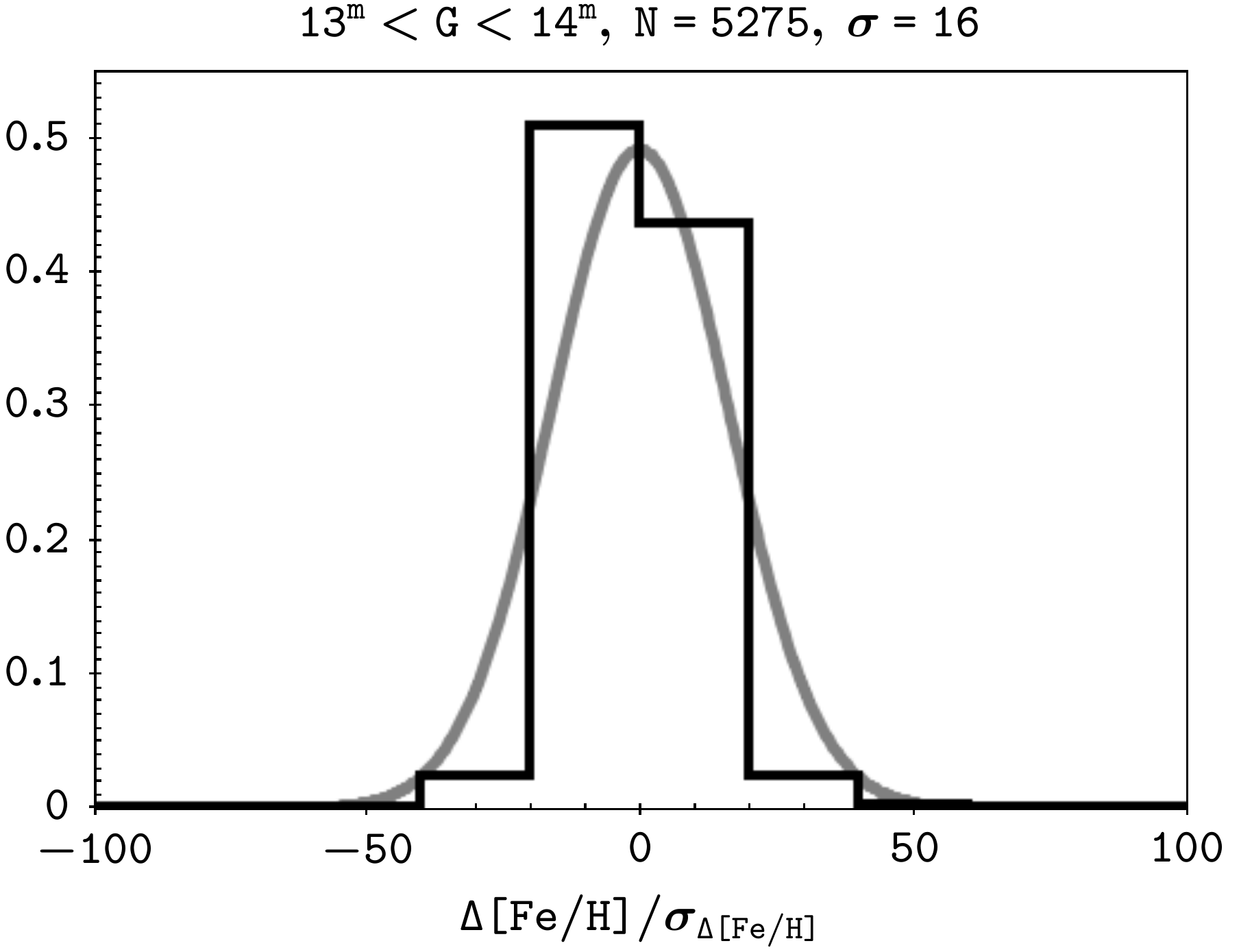}   
             \hfill
         \centering
         \includegraphics[width=7cm]{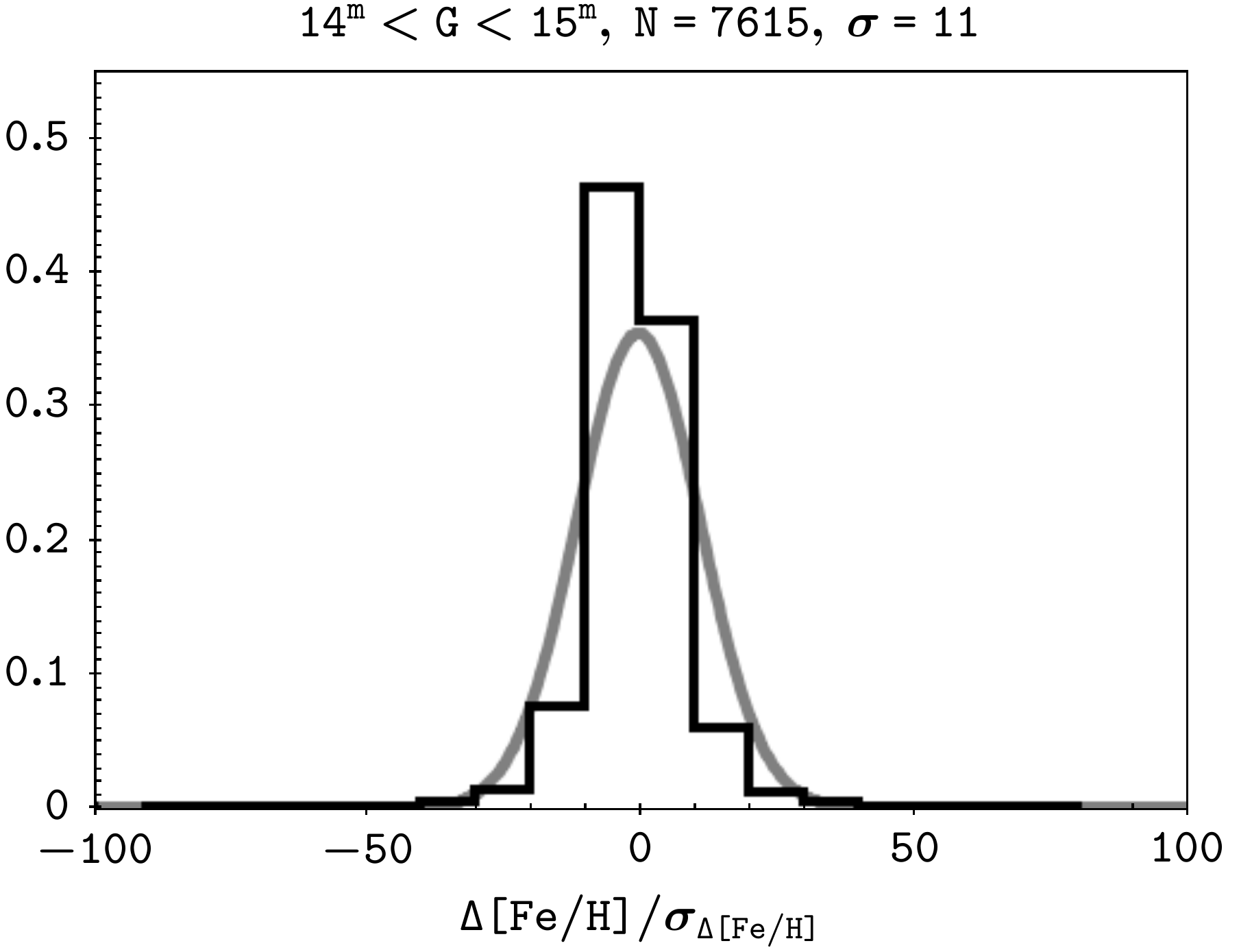}
     \hfill
         \centering
         \includegraphics[width=7cm]{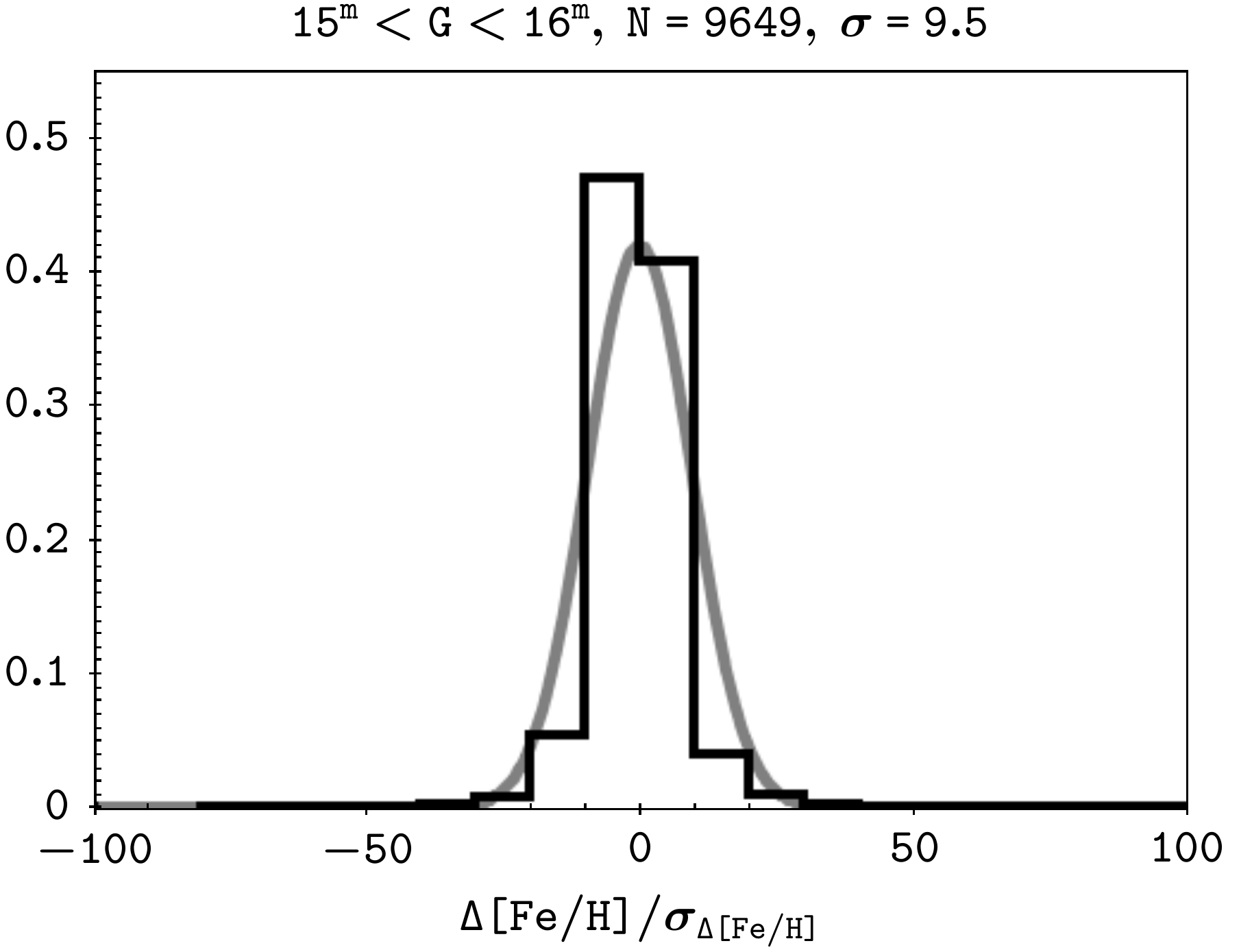}   
    \centering
         \includegraphics[width=7cm]{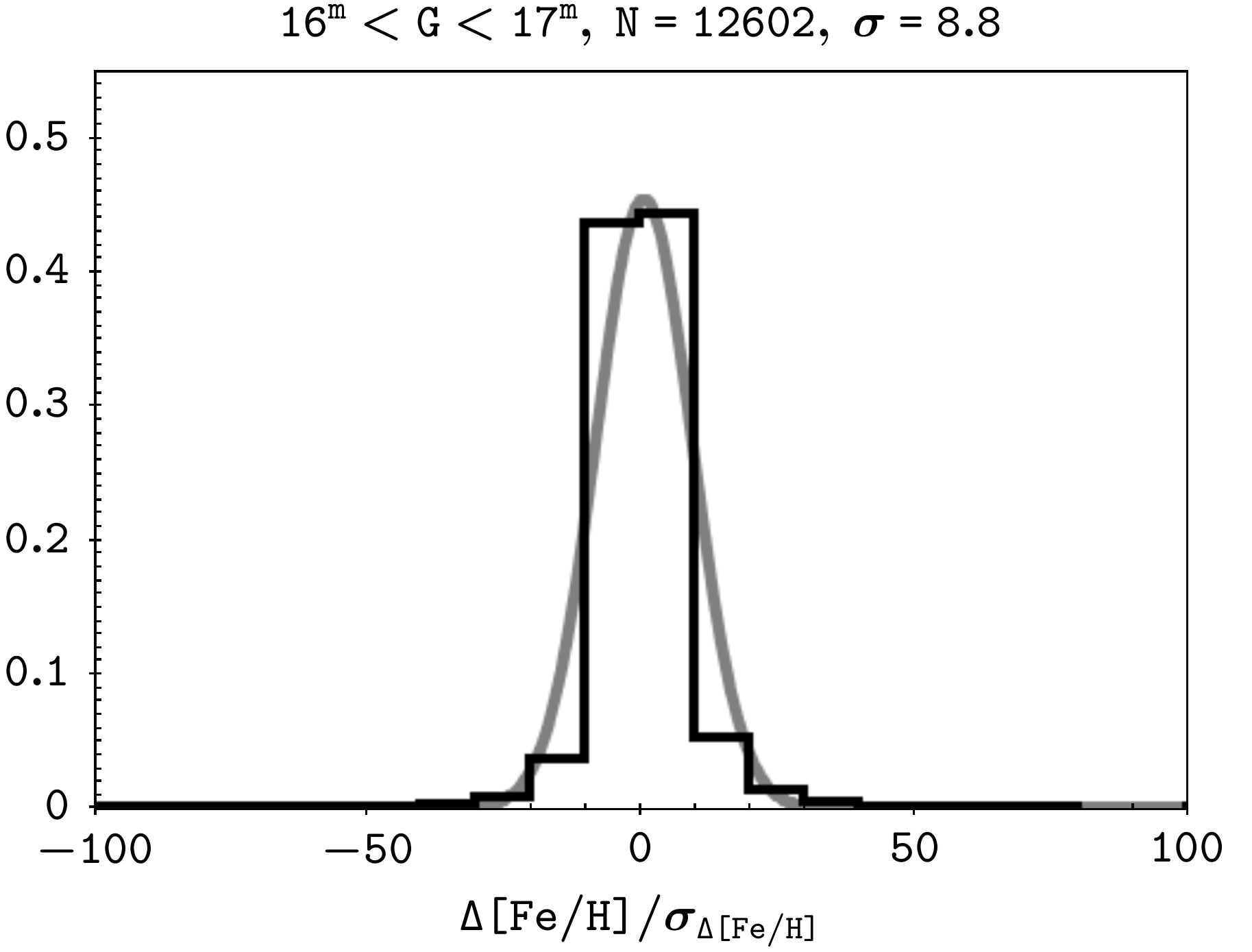}   
             \hfill
         \centering
         \includegraphics[width=7cm]{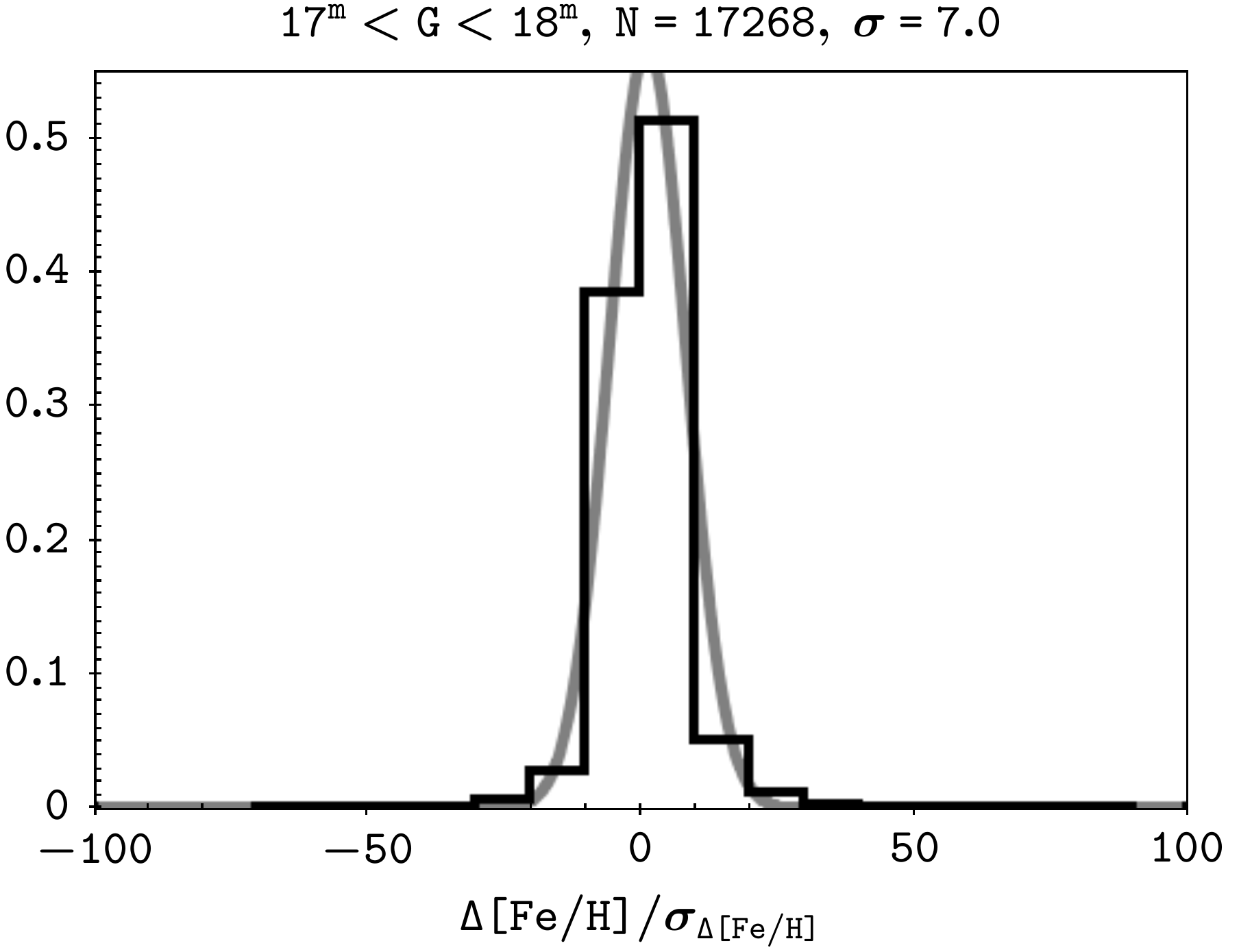}
  
  \caption{Distribution of the metallicity difference between the components of the pair, normalized to the error, for stars in different
ranges of visible magnitude G. A pair is included in the range statistics if both components have a magnitude within the specified
range. Black histograms are observed distributions, gray lines are their approximation by Gaussians. Above each panel is the range
of stellar magnitudes, the number of pairs participating in the statistics, and the standard deviation characterizing the Gaussian
approximation of the distribution. }
  
     \label{FigdFEH}
     \end{figure}

\begin{table}[b]
 \begin{center}
 \caption{Median errors of metallicity [Fe/H] depending
on magnitude}
 \label{tableFEH}
 \medskip
\begin{tabular}[c]{|c|c|c|}
\hline
G, mag& $\sigma_{[Fe/H]}\_{DR3}$ & $\sigma_{Fe/H}\_{exp}$    \\

\hline

<10      & 0.014 & 0.35     \\
10-12     & 0.013 & 0.25   \\
12-13     & 0.014 & 0.17     \\
13-14     & 0.017 & 0.27    \\
14-15     & 0.022 & 0.24   \\
15-16     & 0.029 & 0.28   \\
16-17     & 0.041 & 0.36   \\
17-18     & 0.056 & 0.39   \\
\hline 
\end{tabular}
\end{center}
The median values $\sigma_{[Fe/H]}\_{DR3}$ from Gaia DR3 and the resulting estimates of the expected values of [Fe/H] error $\sigma_{Fe/H}\_{exp}$ are given (asymmetry of the errors is neglected).
\end{table}

A comparison of independently determined characteristics of the components allows us to estimate the
scale of the errors in these quantities, similarly to how
it was done in \cite{2021MNRAS.506.2269E} for parallax errors. We have made
such estimates for the errors in radial velocities and
metallicities of stars from Gaia DR3.

Let us consider the distribution of uncertainties
observed for reliable catalog pairs $\Delta RV/\sigma_{\Delta RV}$, where

\begin{equation*}
 \Delta RV=RV1-RV2,   
\end{equation*}
\begin{equation*}
 \sigma_{\Delta RV}=\sqrt{\sigma_{RV1}^2+\sigma_{RV2}^2}   
\end{equation*}

If the radial velocities of the components are identical
and the radial velocity errors $\sigma_{RV1}, \sigma_{RV2}$ are realistic,
then the distribution of $\Delta RV/\sigma_{\Delta RV}$  will be an unbiased
normal distribution with a variance close to $\sigma^2=1$.

At the Fig.~\ref{FigdRV} shows the error-normalized distributions
of the difference in radial velocities of the components
of the pair in different ranges of visible magnitude .
A pair was included in range statistics only if the magnitudes of both components belonged to the corresponding range. The resulting standard deviations of
the approximating Gaussian distributions are indicated above each panel. Their values vary from 1.5 to
4.9. In this case, the characteristic nominal radial
velocity errors depend on the apparent magnitude of
the source.
In addition, as discussed in Section~\ref{res}, for each
magnitude range a difference between the radial
velocities of the components is expected due to the
presence of orbital motion in the system. Upper estimates of such differences are shown in Fig.~\ref{FigVmeanorb}.
We summarize the results obtained in Table~\ref{tableRV},
where we indicate for different ranges of magnitude
the median radial velocity error from Gaia DR3
$\sigma_{RV}\_{DR3}$, the standard deviation of the normalized difference in the radial velocities of the components of
the pair $\sigma_{RV}\_{bin}$, the median estimate of the possible
difference in the radial velocities of the pair associated
with the orbital motion in the binary system $\sigma_{RV}\_{orb}$,
and our resulting estimate of the expected characteristic radial velocity error $\sigma_{RV}\_{exp}$.
A similar study was carried out to estimate metallicities [Fe/H], assuming identical metallicities for the
components of a pair of stars. In the Gaia DR3 catalog, for astrophysical properties determined by Bayesian methods, the lower $b\_[Fe/H]$ and upper $B\_[Fe/H]$ bounds
are given on the range of probable values that are
asymmetrical with respect to the best estimate. Instead
of the root-mean-square metallicity error, to normalize the observed distribution of differences in [Fe/H]
of the components $\Delta [Fe/H]/\sigma_{\Delta [Fe/H]}$, we (generally speaking, not quite correctly) used the quantity

\begin{equation*}
 \sigma_{\Delta [Fe/H]}=\sqrt{((B\_[Fe/H]_1-b\_[Fe/H]_1)/2)^2+((B\_[Fe/H]_2-b\_[Fe/H]_2)/2)^2}   
\end{equation*}

The distributions of normalized metallicity deviations
are shown in Fig.~\ref{FigdFEH}. It was found that the indicated
ranges of metallicity values in the catalog are apparently underestimated by an order of magnitude.
Median values of $\sigma_{[Fe/H]}\_{DR3}$ from Gaia DR3 and 
the resulting estimates of the expected error values of
[Fe/H] $\sigma_{Fe/H}\_{exp}$ (without considering asymmetry)
are given in Table~\ref{tableFEH}.  

Note also that a study of the statistical agreement of
the characteristics of the components of reliable pairs
(with $R\_chance\_align<0.1$) was carried out in the
original article by \cite{2021MNRAS.506.2269E} using LAMOST data \citep{2012RAA....12.1197C}. For
LAMOST radial velocities, the agreement was found
to be in good agreement with the errors, and the
metallicity errors reported in the catalog were found to
be underestimated.

\section {Conclusions}
\label{concl}
The catalog of wide binary stars \cite{2021MNRAS.506.2269E}, created on the
basis of Gaia EDR3 data and representing the most
extensive list of binary stars, with a high degree of
homogeneity and completeness was analyzed. This list
represents promising material for studying the characteristics of the population of wide binary stars in the
Galactic field.
It was shown that the spatial heterogeneity of the
catalog reflects the Gaia scanning law. This is
explained by the fact that in the areas of the sky
scanned the greatest number of times, the maximum
limiting magnitude and angular resolution of the catalog are achieved.
The change in the spatial density of binary stars in
the catalog with increasing distance from the Sun has
been studied. A model (expected) distribution was
constructed; it is shown by comparing the observed
distribution with it that the catalog contains approximately 2.5 times fewer binary stars than would be
expected in the absence of spatial incompleteness
(when the model distribution is normalized to the
density of binary stars near the Sun). It is confirmed
that the radius of spatial completeness of the catalog is
on average close to 200 pc. It is shown that the radius
of spatial completeness depends on the absolute magnitude of the main component. It increases for more
absolutely bright stars, but at the same time for them
the inner radius of the visibility zone also increases.
The spatial density of binary stars in the catalog is
almost independent of the difference in the magnitudes of the components, but in the near-solar and
most distant regions the catalog is incomplete for pairs
with large brightness differences. The spatial completeness of the catalog depends significantly on the
physical distance between the components. The
incompleteness of the catalog for pairs with a distance
between components less than 100 AU starts already at
a distance of 25 pc from the Sun.

Comparison of the characteristics of components
of the same pair independently determined within the
Gaia DR3 catalog allowed us to study how the probability of a non-random combination of components is
related to the similarity of their characteristics. For
radial velocities, it was found that the median difference magnitude for reliable pairs is 5–6 times less than
the same value for random pairs. Qualitative agreement is also observed for [Fe/H] metallicity estimates
and, to a lesser extent, for absorption $A_G$ estimates. For
estimates of the ages of stars as a complete ensemble,
which is dominated by main sequence stars, no agreement was found, which indicates a large uncertainty in
this value. At the same time, the median agreement of
age estimates for pairs with evolved components is
much better and is about $25\%$.
Using the parameters of the pair components from
Gaia DR3, an independent estimate of the uncertainties in the radial velocities and metallicities depending
on the apparent magnitude of the sources was performed. When comparing the radial velocities of the
components, their probable difference due to the presence of orbital motion in the binary system was considered. Estimates of the expected difference are made
depending on the distance to the system and on the
apparent magnitude of the main component. Estimates of the probable median values of errors in the
radial velocities and metallicities of Gaia DR3 sources
are proposed. Depending on the apparent magnitude,
they exceed the median error values given in the catalog: for radial velocities by 1.5–3 times, for metallicities [Fe/H] by 7–25 times.

\bigskip

      The work uses data from {\it Gaia} space mission of the
European Space Agency (ESA), (\url{https://www.cosmos.esa.int/gaia}), processed by the {\it Gaia} Data Processing and Analysis Consortium (DPAC, \url{https://www.cosmos.esa.int/web/gaia/dpac/consortium}). An interactive graphical visualizer
and analyzer for tabular data TOPCAT  \citep{2005ASPC..347...29T} was used. 

      The author is grateful to the anonymous
reviewer for stimulating comments and recommendations
that made it possible to improve the work.

\bigskip

The author declares no conflicts of interest.

\bibliographystyle{aspb1}
\bibliography{fieldbin}

\begin{thebibliography}{27}
\providecommand{\natexlab}[1]{#1}

\bibitem[{Bailer-Jones} et~al.(2021)]{2021AJ....161..147B}
C.~A.~L. {Bailer-Jones}, J.~{Rybizki}, M.~{Fouesneau}, et~al., \aj \textbf{161}~(3), 147 (2021).

\bibitem[{Bate}(2015)]{2015ASPC..496...37B}
M.~R. {Bate}, in S.~M. {Rucinski}, G.~{Torres}, and M.~{Zejda} (eds.), \emph{Living Together: Planets, Host Stars and Binaries}, \emph{Astronomical Society of the Pacific Conference Series}, vol. 496, p.~37 (2015).

\bibitem[{Bovy}(2017)]{2017MNRAS.470.1360B}
J.~{Bovy}, \mnras \textbf{470}~(2), 1360 (2017).

\bibitem[{Chulkov} and {Malkov}(2022)]{2022MNRAS.517.2925C}
D.~{Chulkov} and O.~{Malkov}, \mnras \textbf{517}~(2), 2925 (2022).

\bibitem[{Creevey} et~al.(2022)]{2022arXiv220605864C}
O.~L. {Creevey}, R.~{Sordo}, F.~{Pailler}, et~al., arXiv e-prints arXiv:2206.05864 (2022).

\bibitem[{Cui} et~al.(2012)]{2012RAA....12.1197C}
X.-Q. {Cui}, Y.-H. {Zhao}, Y.-Q. {Chu}, et~al., Research in Astronomy and Astrophysics \textbf{12}~(9), 1197 (2012).

\bibitem[{Duch{\^e}ne} and {Kraus}(2013)]{2013ARA&A..51..269D}
G.~{Duch{\^e}ne} and A.~{Kraus}, \araa \textbf{51}~(1), 269 (2013).

\bibitem[{El-Badry} et~al.(2021)]{2021MNRAS.506.2269E}
K.~{El-Badry}, H.-W. {Rix}, and T.~M. {Heintz}, \mnras \textbf{506}~(2), 2269 (2021).

\bibitem[{El-Badry} et~al.(2019)]{Badry}
K.~{El-Badry}, H.-W. {Rix}, H.~{Tian}, et~al., Mon.~Not.~R.~Astron.~Soc \textbf{489}~(4), 5822 (2019).

\bibitem[{Foley} et~al.(2022)]{2022AAS...24033303F}
M.~{Foley}, A.~{Goodman}, C.~{Zucker}, et~al., in \emph{American Astronomical Society Meeting Abstracts}, \emph{American Astronomical Society Meeting Abstracts}, vol.~54, p. 333.03 (2022).

\bibitem[{Gaia Collaboration} et~al.(2021)]{2021A&A...649A...1G}
{Gaia Collaboration}, A.~G.~A. {Brown}, A.~{Vallenari}, et~al., \aap \textbf{649}, A1 (2021).

\bibitem[{Gaia Collaboration} et~al.(2016)]{2016A&A...595A...1G}
{Gaia Collaboration}, T.~{Prusti}, J.~H.~J. {de Bruijne}, et~al., \aap \textbf{595}, A1 (2016).

\bibitem[{Hartkopf} et~al.(2001)]{2001AJ....122.3472H}
W.~I. {Hartkopf}, B.~D. {Mason}, and C.~E. {Worley}, \aj \textbf{122}~(6), 3472 (2001).

\bibitem[{Katz} et~al.(2022)]{2022arXiv220605902K}
D.~{Katz}, P.~{Sartoretti}, A.~{Guerrier}, et~al., arXiv e-prints arXiv:2206.05902 (2022).

\bibitem[{Kharchenko} et~al.(2016)]{2016A&A...585A.101K}
N.~V. {Kharchenko}, A.~E. {Piskunov}, E.~{Schilbach}, et~al., \aap \textbf{585}, A101 (2016).

\bibitem[Kovaleva et~al.(2021)]{Kovaleva2021125}
D.~Kovaleva, O.~Malkov, S.~Sapozhnikov, et~al., Communications in Computer and Information Science \textbf{1427}, 125 – 133 (2021).

\bibitem[{Marks} and {Kroupa}(2012)]{2012A&A...543A...8M}
M.~{Marks} and P.~{Kroupa}, \aap \textbf{543}, A8 (2012).

\bibitem[{Marks} et~al.(2022)]{2022A&A...659A..96M}
M.~{Marks}, P.~{Kroupa}, and J.~{Dabringhausen}, \aap \textbf{659}, A96 (2022).

\bibitem[{Moe} and {Di Stefano}(2017)]{2017ApJS..230...15M}
M.~{Moe} and R.~{Di Stefano}, \apjs \textbf{230}~(2), 15 (2017).

\bibitem[{Pecaut} and {Mamajek}(2013)]{2013ApJS..208....9P}
M.~J. {Pecaut} and E.~E. {Mamajek}, \apjs \textbf{208}~(1), 9 (2013).

\bibitem[{Popova} et~al.(1982)]{1982Ap&SS..88...55P}
E.~I. {Popova}, A.~V. {Tutukov}, and L.~R. {Yungelson}, \apss \textbf{88}~(1), 55 (1982).

\bibitem[{Pourbaix} et~al.(2022)]{2022gdr3.reptE...7P}
D.~{Pourbaix}, F.~{Arenou}, P.~{Gavras}, et~al., {Gaia DR3 documentation Chapter 7: Non-single stars}, Gaia DR3 documentation, European Space Agency; Gaia Data Processing and Analysis Consortium. https://gea.esac.esa.int/archive/documentation/GDR3/index.html</A>, id. 7 (2022).

\bibitem[Rozner and Perets(2023)]{rozner2023born}
M.~Rozner and H.~B. Perets, Born to be wide: the distribution of wide binaries in the field and soft binaries in clusters (2023).

\bibitem[{Sapozhnikov} et~al.(2020)]{our}
S.~A. {Sapozhnikov}, D.~A. {Kovaleva}, O.~Y. {Malkov}, and A.~Y. {Sytov}, Astronomy Reports \textbf{64}~(9), 756 (2020).

\bibitem[{Taylor}(2005)]{2005ASPC..347...29T}
M.~B. {Taylor}, in P.~{Shopbell}, M.~{Britton}, and R.~{Ebert} (eds.), \emph{Astronomical Data Analysis Software and Systems XIV}, \emph{Astronomical Society of the Pacific Conference Series}, vol. 347, p.~29 (2005).

\bibitem[{Tokovinin}(2000)]{2000A&A...360..997T}
A.~A. {Tokovinin}, \aap \textbf{360}, 997 (2000).

\bibitem[{Vereshchagin} et~al.(1988)]{1988Ap&SS.142..245V}
S.~{Vereshchagin}, A.~{Tutukov}, L.~{Iungelson}, et~al., \apss \textbf{142}~(1-2), 245 (1988).

\end{thebibliography}

\end{document}